%% file: arXiv.tex
\documentclass[a4paper,fleqn]{cas-sc}

\usepackage[numbers,sort&compress]{natbib}
\usepackage[utf8]{inputenc}
\usepackage{booktabs}
\usepackage{amsmath,amssymb,amsfonts,mathtools}
\usepackage{array}
\usepackage{braket}
\usepackage{graphicx}
\graphicspath{{./figures/}}
\usepackage{algorithm}
\usepackage{algpseudocode}
\usepackage{csquotes}
\usepackage{hhline}
\usepackage{listings}
\usepackage{xcolor}
\usepackage{subcaption}
\usepackage{dcolumn}
\usepackage{tabularx}
\usepackage{longtable}
\usepackage{tikz}
\usepackage{hyperref}
\hypersetup{colorlinks=true,linktoc=all,linkcolor=black,citecolor=magenta,urlcolor=blue}

\newcolumntype{C}{>{\centering\arraybackslash}X}
\begin{document}
\let\WriteBookmarks\relax
\def\floatpagepagefraction{1}
\def\textpagefraction{.001}

\shorttitle{QSVM-RQNN for Condition Monitoring and Fault Classification}
\shortauthors{Patel et al.}
\title[mode=title]{QSVM-RQNN: Low-Qubit Recurrent Quantum Similarity Learning for Condition Monitoring and Fault Classification}

\author[1]{Amit S. Patel}
\ead{aspatel.mh@ddu.ac.in}
\author[2]{Himanshu R. Patel}
\ead{himanshupatel.ic@ddu.ac.in}
\author[3,4]{Bikash K. Behera}
\ead{bikas.riki@gmail.com}
\author[5]{Pranav Mehta}
\ead{pranavmehta.mech@ddu.ac.in}

\author[6,7]{David Bassir}
\cormark[1]
\ead{david.bassir@ens-paris-saclay.fr}

\author[8,9]{Ghanshyam G. Tejani}
\ead{gtejani@saturn.yzu.edu.tw}

\affiliation[1]{organization={Department of Mechanical Engineering, Dharmsinh Desai University},city={Nadiad},state={Gujarat},country={India}}
\affiliation[2]{organization={Department of Instrumentation \& Control Engineering, Dharmsinh Desai University},city={Nadiad},state={Gujarat},country={India}}
\affiliation[3]{organization={Bikash's Quantum (OPC) Pvt. Ltd.},city={Mohanpur},postcode={741246},state={West Bengal},country={India}}
\affiliation[4]{organization={Universit\`a degli Studi di Cagliari},addressline={Via Is Mirrions},city={Cagliari},postcode={09123},country={Italy}}

\affiliation[5]{organization={Department of Mechanical Engineering, Dharmsinh Desai University},
            city={Nadiad},
            state={Gujarat},
            postcode={387001},
            country={India}}

\affiliation[6]{organization={Smart Structural Health Monitoring and Control Laboratory, DGUT-CNAM, Dongguan University of Technology},
            city={Dongguan},
            country={China}}

\affiliation[7]{organization={ENS-Paris-Saclay University, Centre Borelli, UMR CNRS 9010},
            city={Gif-sur-Yvette},
            postcode={91190},
            country={France}}

\affiliation[8]{organization={Department of Research Analytics, Saveetha Dental College and Hospitals, Saveetha Institute of Medical and Technical Sciences, Saveetha University},
            city={Chennai},
            postcode={600077},
            country={India}}

\affiliation[9]{organization={Department of Industrial Engineering and Management, Yuan Ze University},
            city={Taoyuan},
            postcode={320315},
            country={Taiwan}}
\cortext[1]{Corresponding author}

\begin{abstract}
In the Noisy Intermediate-Scale Quantum (NISQ) era, limited qubit availability and hardware noise remain major challenges for deploying quantum machine learning (QML) algorithms in practical applications. Existing quantum neural network (QNN) and quantum convolutional neural network (QCNN) architectures often require increasing quantum resources as the input feature dimension grows, limiting their scalability on near-term quantum devices. To address this challenge, we propose QSVM-RQNN, a low-qubit recurrent quantum machine learning framework that integrates centroid-based Quantum Support Vector Machine (QSVM) similarity learning with Recurrent Quantum Neural Networks (RQNNs) for intelligent fault classification. The proposed framework first reduces the input feature space using principal component analysis (PCA), partitions the reduced representation into sequential timesteps, and processes them through a compact three-qubit recurrent quantum architecture with shared parameters. Two complementary variants are developed: QSVM-RQNN-V1 performs class-conditioned joint quantum encoding of the input and class-centroid segments, whereas QSVM-RQNN-V2 performs recurrent learning over timestep-wise quantum similarity representations. This design enables expressive sequence modeling while maintaining low quantum resource requirements and computational efficiency. Experimental evaluation on multiple fault diagnosis datasets demonstrates that the proposed QSVM-RQNN framework consistently achieves competitive and, in several cases, state-of-the-art performance compared with standalone QSVM, QNN, QCNN, QSVM-QNN, QSVM-QCNN, and RQNN models. In particular, the proposed architectures achieve superior performance-efficiency trade-offs, improved recall, and enhanced fault detection on highly imbalanced datasets. These results demonstrate that integrating centroid-based quantum similarity learning with low-qubit recurrent quantum representation learning provides an effective and scalable solution for condition monitoring and fault classification on resource-constrained NISQ devices.
\end{abstract}


\maketitle
\section{Introduction}\label{Sec1}

\subsection{Context and Motivation}

The rapid advancement of Industry 4.0 has accelerated the deployment of intelligent sensing technologies and data-driven monitoring systems across manufacturing, transportation, energy, aerospace, and smart infrastructure. Modern industrial assets continuously generate large volumes of heterogeneous sensor measurements, vibration signals, acoustic emissions, thermal images, and visual inspection data, enabling continuous assessment of equipment health and operational reliability. Accurate condition monitoring and early fault diagnosis have become essential for reducing unexpected downtime, improving system reliability, minimizing maintenance costs, and ensuring operational safety. Consequently, intelligent fault diagnosis has become an active research area at the intersection of machine learning, signal processing, and industrial artificial intelligence \cite{lee2015cyber,zhang2019deep,yan2020industrial}.

Recent advances in machine learning and deep learning have significantly improved the capability of automated fault diagnosis systems. Conventional approaches based on support vector machines (SVM), random forests (RF), convolutional neural networks (CNNs), recurrent neural networks (RNNs), long short-term memory (LSTM) networks, and transformer-based architectures have demonstrated remarkable performance in bearing fault diagnosis, photovoltaic defect inspection, rotating machinery monitoring, and predictive maintenance applications \cite{lecun2015deep,goodfellow2016deep,wang2021deep}. These methods are capable of learning complex nonlinear relationships directly from raw measurements and have substantially reduced the need for handcrafted feature engineering. Nevertheless, their performance often depends on large quantities of labeled training data and computationally intensive optimization, while highly imbalanced fault datasets remain a persistent challenge in practical industrial environments.

Quantum computing has emerged as a promising computational paradigm capable of exploiting quantum superposition, entanglement, and interference to perform information processing fundamentally differently from classical computing \cite{nielsen2010quantum,preskill2018quantum}. Motivated by recent progress in noisy intermediate-scale quantum (NISQ) hardware, quantum machine learning (QML) has attracted considerable attention as a potential approach for improving learning efficiency and representation capability using parameterized quantum circuits \cite{schuld2015introduction,biamonte2017quantum,schuld2021machine,cerezo2021variational}. Quantum classifiers, variational quantum circuits, quantum neural networks (QNNs), quantum convolutional neural networks (QCNNs), and quantum kernel methods have been successfully investigated for classification, regression, optimization, image analysis, and industrial fault diagnosis \cite{farhi2018classification,cong2019quantum,havlivcek2019supervised,schuld2019quantum}.

Despite these encouraging developments, practical deployment of QML remains constrained by the limited resources of current NISQ devices. Existing quantum learning architectures frequently require increasing numbers of qubits and deeper quantum circuits as the input feature dimension grows, leading to higher hardware requirements, increased circuit depth, and greater sensitivity to quantum noise and decoherence \cite{preskill2018quantum,cerezo2021variational}. These limitations become particularly significant in industrial condition monitoring applications, where high-dimensional feature representations and limited fault samples often coexist. Consequently, developing computationally efficient, low-qubit quantum learning architectures capable of maintaining high classification performance while operating within the constraints of near-term quantum hardware has become an important research direction.

\subsection{Research Gap Analysis}

Although QML has demonstrated significant potential for classification and optimization, existing quantum learning architectures still face several challenges that limit their practical deployment for industrial condition monitoring on NISQ devices. Recent studies have investigated Quantum Support Vector Machines (QSVMs), QNNs, QCNNs, Variational Quantum Classifiers (VQCs), and recurrent quantum architectures for pattern recognition and fault diagnosis \cite{cerezo2021variational,farhi2018classification,cong2019quantum,havlivcek2019supervised,schuld2019quantum}. While these approaches have achieved encouraging results, they exhibit complementary strengths and limitations when applied to high-dimensional industrial datasets. QSVMs exploit quantum feature maps to improve class separability by embedding classical data into high-dimensional Hilbert spaces \cite{havlivcek2019supervised,schuld2019quantum}. However, conventional QSVMs require the construction of quantum kernel matrices whose computational cost increases with the number of training samples, while the entire feature vector is typically encoded into a single quantum state, resulting in increasing circuit width and encoding complexity for high-dimensional data. These factors limit their scalability for practical industrial applications.

Trainable quantum models, including QNNs, QCNNs, and VQCs, overcome some limitations of kernel-based methods by learning adaptive quantum representations directly from data \cite{cerezo2021variational,cong2019quantum,schuld2020circuit}. Nevertheless, accurately modeling complex feature interactions often requires wider or deeper parameterized quantum circuits, increasing optimization difficulty and susceptibility to quantum noise, decoherence, and barren plateau phenomena on NISQ hardware \cite{cerezo2021variational,mcclean2018barren}. Recurrent Quantum Neural Networks (RQNNs) improve parameter efficiency through sequential processing and parameter sharing, making them attractive for low-qubit implementations \cite{beer2020training,abbas2021power}. However, existing RQNN-based models generally learn directly from sequentially encoded features without explicitly incorporating similarity-guided learning, limiting their ability to exploit discriminative class relationships. In addition, industrial fault diagnosis datasets are frequently highly imbalanced, yet relatively few quantum learning frameworks integrate effective data balancing strategies with low-qubit quantum representation learning.

These limitations reveal several important research gaps. Existing quantum classifiers rarely achieve low-qubit implementation, computational efficiency, and strong classification performance simultaneously. Moreover, quantum similarity learning and recurrent quantum representation learning have largely been investigated independently despite their complementary strengths, while the integration of class balancing with compact quantum learning architectures remains largely unexplored. Motivated by these challenges, this work proposes the QSVM-RQNN framework, which unifies centroid-based quantum similarity learning, recurrent quantum representation learning, and QSMOTE-based data balancing within a compact three-qubit architecture for efficient and robust industrial fault classification.

\begin{figure}[pos=htbp]
\centering
\includegraphics[width=\textwidth]{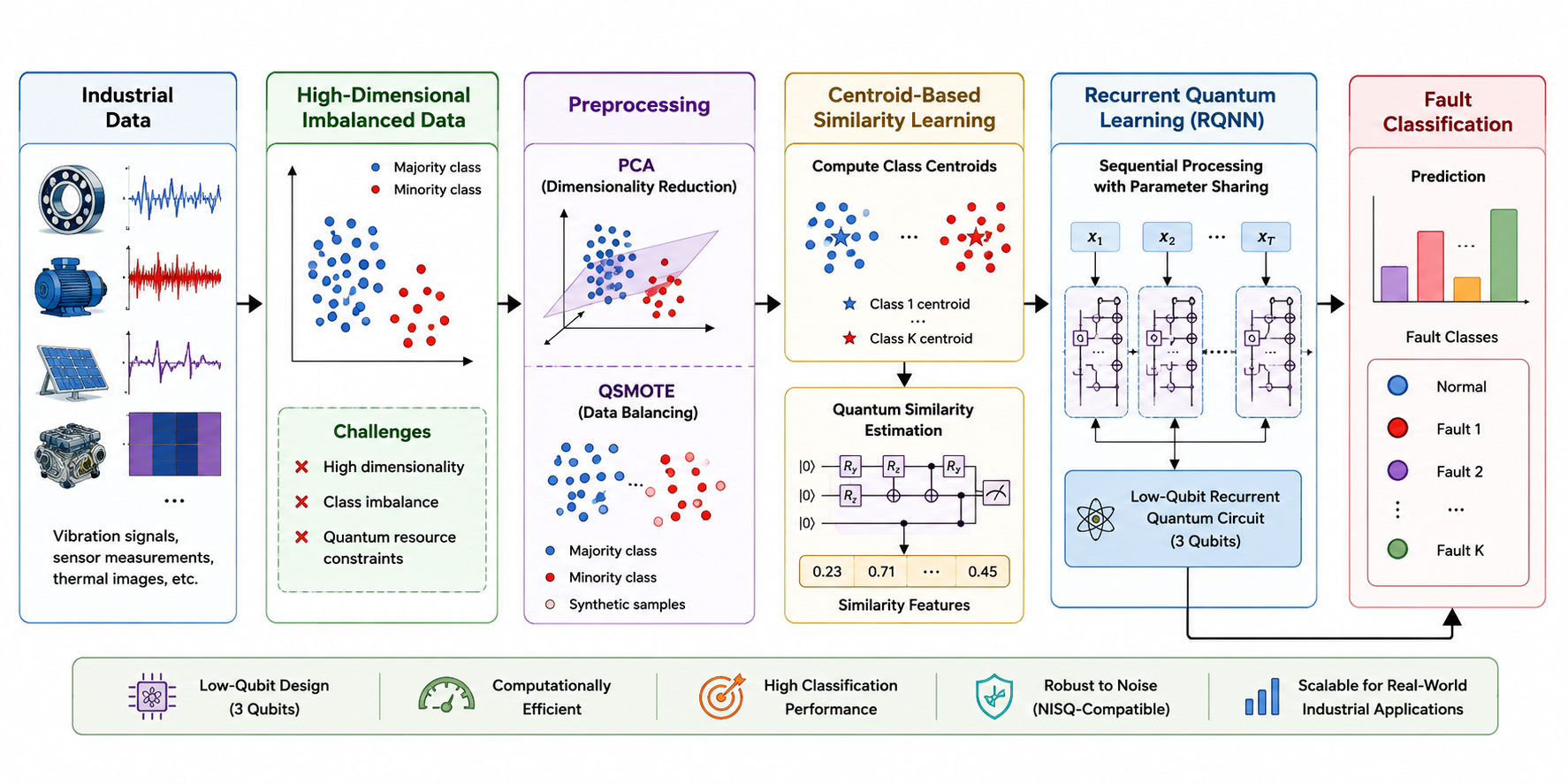}
\caption{Conceptual overview of the proposed QSVM-RQNN framework. High-dimensional industrial data are first transformed into compact representations through PCA and balanced using QSMOTE. Centroid-based quantum similarity learning subsequently generates discriminative representations that are processed by a low-qubit recurrent quantum network for efficient fault classification on NISQ devices.}
\label{fig:overall_schematic}
\end{figure}

\subsection{Proposed QSVM-RQNN Framework}

Motivated by the above research gaps, this work proposes QSVM-RQNN, a low-qubit hybrid QML framework that integrates centroid-based QSVM similarity learning with RQNNs for efficient industrial fault classification. Rather than encoding an entire high-dimensional feature vector into a large quantum circuit, the proposed framework progressively transforms the data into compact sequential quantum representations, making it suitable for resource-constrained NISQ devices. Figure~\ref{fig:overall_schematic} illustrates the overall workflow of the proposed framework. The input measurements are first preprocessed using Principal Component Analysis (PCA) to obtain a compact feature representation, followed by Quantum Synthetic Minority Over-sampling Technique (QSMOTE) \cite{Mohanty2025QuantumSMOTE} to alleviate class imbalance. Representative class centroids are then constructed to enable centroid-based quantum similarity learning, providing discriminative similarity information without explicitly constructing large quantum kernel matrices.

The resulting similarity representations are processed using a compact recurrent quantum architecture with shared parameters across all timesteps, enabling expressive sequential representation learning while requiring only a fixed three-qubit quantum circuit. To investigate different strategies for incorporating similarity information, two complementary variants are developed. QSVM-RQNN-V1 performs class-conditioned joint quantum encoding of the input features and class centroids prior to recurrent processing, whereas QSVM-RQNN-V2 first computes timestep-wise similarity features before recurrent quantum learning. Although both variants share the same recurrent backbone, they differ in how similarity information is incorporated into the learning process. By integrating dimensionality reduction, quantum data balancing, centroid-based similarity learning, and recurrent quantum representation learning within a unified low-qubit architecture, QSVM-RQNN provides an efficient and scalable solution for intelligent condition monitoring on NISQ devices. The complete mathematical formulation is presented in Section~\ref{SecIII}.

\subsection{Main Contributions}

The main contributions of this work are summarized as follows.

\begin{itemize}

\item We propose \textbf{QSVM-RQNN}, a novel low-qubit hybrid quantum learning framework that integrates centroid-based QSVM similarity learning with RQNNs for efficient condition monitoring and fault classification on NISQ devices.

\item We develop two complementary architectures, \textbf{QSVM-RQNN-V1} and \textbf{QSVM-RQNN-V2}, which incorporate similarity information at different stages of recurrent quantum learning, enabling a systematic investigation of alternative similarity-guided learning strategies.

\item We introduce a computationally efficient pipeline that combines PCA-based dimensionality reduction, QSMOTE-based class balancing, centroid-based quantum similarity learning, and a compact three-qubit recurrent quantum architecture with shared parameters, avoiding explicit quantum kernel matrix construction.

\item We validate the proposed framework on multiple benchmark fault diagnosis datasets. Experimental results demonstrate competitive and, in several cases, state-of-the-art performance compared with QSVM, QNN, QCNN, QSVM-QNN \cite{behera2025qsvmqnn}, QSVM-QCNN, and RQNN models, while improving minority-class fault detection and performance efficiency.

\end{itemize}

\subsection{Organization of the Paper}
The remainder of this paper is organized as follows. Section~\ref{SecII} reviews the relevant background and related work. Section~\ref{SecIII} presents the proposed QSVM-RQNN framework. Section~\ref{SecIV} describes the experimental setup, implementation details and discusses the experimental results. Finally, Section~\ref{SecV} concludes the paper and outlines future research directions.

\section{Background and Related Work}\label{SecII}

\subsection{Quantum Support Vector Machines}

SVMs are among the most successful supervised learning algorithms owing to their strong theoretical foundation, excellent generalization capability, and effectiveness on limited training data \cite{cortes1995support,vapnik1998statistical}. They have been widely applied to industrial condition monitoring tasks such as bearing fault diagnosis, rotating machinery monitoring, structural health monitoring, and defect detection \cite{widodo2007support,shao2011fault}. However, classical SVMs rely on carefully designed kernel functions, whose computational cost increases for large-scale and high-dimensional datasets.

QSVMs extend classical SVMs by mapping data into quantum Hilbert spaces through parameterized quantum feature maps before similarity evaluation \cite{havlivcek2019supervised,rebentrost2014quantum}. The resulting quantum kernels can capture highly nonlinear feature relationships that are difficult to construct classically, making QSVM one of the most widely studied quantum classifiers for NISQ devices \cite{schuld2021machine,schuld2019quantum}. Recent studies have successfully applied QSVMs to image classification, medical diagnosis, predictive maintenance, and industrial fault diagnosis, demonstrating the effectiveness of quantum similarity learning for nonlinear classification tasks \cite{li2022quantum,wang2024quantum,diedrich2024quantum}.

Despite these advances, conventional QSVMs still face several challenges. The entire feature vector is typically encoded into a single quantum state, leading to increasing circuit width and encoding complexity as the input dimensionality grows. In addition, constructing and evaluating quantum kernel matrices becomes increasingly expensive with larger training datasets, limiting scalability for practical industrial applications. Furthermore, QSVMs primarily perform similarity-based classification without explicitly learning adaptive feature representations, motivating the development of trainable quantum neural architectures capable of more expressive representation learning.

\subsection{Quantum Neural Networks and Quantum Convolutional Neural Networks}

The rapid development of variational quantum algorithms has established QNNs as a prominent paradigm for supervised quantum learning. Unlike quantum kernel methods, QNNs employ parameterized quantum circuits whose trainable gate parameters are optimized through classical algorithms, enabling end-to-end representation learning directly from data \cite{cerezo2021variational,farhi2018classification,schuld2020circuit}. By exploiting quantum superposition and entanglement, QNNs provide flexible nonlinear models compatible with NISQ hardware. QNNs have been successfully applied to classification, regression, reinforcement learning (RL), and industrial fault diagnosis \cite{beer2020training,abbas2021power}. Compared with QSVMs, they avoid explicit kernel matrix construction by directly optimizing parameterized quantum circuits to learn task-specific feature representations.

Inspired by classical convolutional neural networks, QCNNs introduce quantum convolution and pooling operations for hierarchical feature extraction \cite{cong2019quantum}. QCNNs have demonstrated promising performance in quantum phase recognition, image classification, and fault diagnosis through efficient multi-scale quantum representation learning. Despite these advantages, existing QNN and QCNN architectures often require wider quantum circuits or deeper variational layers as the input dimensionality increases, leading to larger optimization spaces and greater susceptibility to hardware noise, decoherence, and barren plateau phenomena on NISQ devices \cite{cerezo2021variational,mcclean2018barren}. These limitations reduce their scalability for high-dimensional industrial condition monitoring tasks. Consequently, although QNNs and QCNNs provide powerful trainable quantum representations, more parameter-efficient architectures are needed. One promising direction is recurrent quantum learning, where parameter sharing enables compact sequential representation learning with reduced quantum resource requirements.

\subsection{Recurrent Quantum Neural Networks}

Motivated by the success of RNNs and LSTM networks, RQNNs extend sequential learning to quantum computing by repeatedly applying a parameterized quantum circuit with shared trainable parameters across multiple timesteps \cite{beer2020training,abbas2021power}. Compared with feedforward quantum models, RQNNs improve parameter efficiency and enable high-dimensional feature vectors to be processed sequentially using a fixed-size quantum register, making them well suited for NISQ devices. Recent studies have investigated RQNNs for quantum sequence learning, time-series prediction, quantum reservoir computing, temporal signal processing, and RL \cite{beer2020training,nakajima2021quantum,fujii2023quantum}. These approaches demonstrate that recurrent quantum processing can effectively capture sequential dependencies while maintaining compact quantum architectures suitable for limited qubit resources.

Despite these advances, existing RQNNs primarily learn directly from sequentially encoded input features without explicitly incorporating similarity-guided learning or prototype-based class information. Consequently, discriminative representation learning and decision boundary construction are performed simultaneously, potentially increasing the learning difficulty. Furthermore, the integration of recurrent quantum learning with quantum similarity estimation and class-balancing strategies remains largely unexplored for industrial fault diagnosis. These limitations motivate the development of hybrid quantum architectures that combine quantum similarity learning with recurrent quantum representation learning, enabling more discriminative and resource-efficient fault classification on NISQ devices.

\subsection{Hybrid Quantum Learning Frameworks}

Recognizing the complementary strengths of quantum kernel methods and trainable QNNs, recent studies have proposed hybrid quantum learning frameworks that integrate multiple quantum paradigms within a unified model. These approaches combine the similarity learning capability of QSVMs with the adaptive representation learning of variational quantum circuits, improving flexibility for tasks such as image classification, pattern recognition, anomaly detection, and industrial fault diagnosis. Representative frameworks include QSVM-QNN \cite{behera2025qsvmqnn} and QSVM-QCNN architectures, where quantum similarity estimation is combined with trainable quantum classifiers to improve nonlinear feature learning. More recently, RL-inspired quantum models and other hybrid variational architectures have also been explored to enhance adaptive representation learning and decision making.

Despite these advances, several challenges remain. Most existing hybrid models employ feedforward variational circuits after similarity learning, resulting in increasing circuit complexity as network depth grows. Moreover, recurrent quantum learning, which enables parameter sharing and sequential feature extraction with fewer trainable parameters, has received comparatively little attention. Existing hybrid approaches also rarely address class imbalance or the efficient processing of high-dimensional industrial data on resource-constrained NISQ hardware. These limitations motivate the development of more unified quantum learning frameworks that combine quantum similarity learning, recurrent quantum representation learning, and robust data preprocessing within a compact architecture. The proposed QSVM-RQNN framework addresses this gap by integrating centroid-based quantum similarity learning, recurrent quantum neural networks with shared parameters, PCA-based feature reduction, and QSMOTE-based class balancing into a unified three-qubit architecture for industrial condition monitoring and fault classification.

\subsection{Summary and Research Positioning}

The above review highlights the complementary strengths of existing quantum learning paradigms. QSVMs provide effective similarity learning but lack adaptive representation learning, whereas QNNs and QCNNs learn expressive quantum representations at the expense of increased circuit complexity. RQNNs improve parameter efficiency through sequential processing, while hybrid quantum frameworks combine multiple paradigms to enhance learning flexibility. Despite these advances, important challenges remain. Existing approaches rarely integrate quantum similarity learning, recurrent representation learning, and class balancing within a unified low-qubit framework, limiting their scalability for high-dimensional and imbalanced industrial datasets. Table~\ref{tab:related_comparison} summarizes the characteristics of representative quantum learning approaches. As shown, the proposed QSVM-RQNN framework uniquely combines centroid-based quantum similarity learning, recurrent quantum representation learning with shared parameters, QSMOTE-based class balancing, and a compact three-qubit architecture within a unified framework for efficient industrial condition monitoring and fault classification. The complete methodology is presented in the following section.

\begin{table}[pos=htbp]
\centering
\caption{Comparison of representative quantum machine learning approaches for condition monitoring and fault classification.}
\label{tab:related_comparison}
\renewcommand{\arraystretch}{1.2}

\begin{tabular}{lcccccc}
\hline
\textbf{Method} &
\shortstack{\textbf{Similarity}\\\textbf{Learning}} &
\shortstack{\textbf{Recurrent}\\\textbf{Learning}} &
\shortstack{\textbf{Low-}\\\textbf{Qubit}} &
\shortstack{\textbf{Class}\\\textbf{Balancing}} &
\shortstack{\textbf{Industrial}\\\textbf{Applications}} &
\shortstack{\textbf{Trainable}\\\textbf{Representation}} \\
\hline

QSVM &
\checkmark &
$\times$ &
$\times$ &
$\times$ &
\checkmark &
$\times$ \\

QNN &
$\times$ &
$\times$ &
$\times$ &
$\times$ &
\checkmark &
\checkmark \\

QCNN &
$\times$ &
$\times$ &
$\times$ &
$\times$ &
\checkmark &
\checkmark \\

RQNN &
$\times$ &
\checkmark &
\checkmark &
$\times$ &
Limited &
\checkmark \\

QSVM-QNN &
\checkmark &
$\times$ &
Partial &
$\times$ &
\checkmark &
\checkmark \\

QSVM-QCNN &
\checkmark &
$\times$ &
Partial &
$\times$ &
\checkmark &
\checkmark \\

Hybrid Quantum Models &
Partial &
Partial &
Partial &
$\times$ &
Limited &
\checkmark \\

\textbf{QSVM-RQNN (Proposed)} &
\textbf{\checkmark} &
\textbf{\checkmark} &
\textbf{\checkmark} &
\textbf{\checkmark (QSMOTE)} &
\textbf{\checkmark} &
\textbf{\checkmark} \\

\hline
\end{tabular}
\end{table}

\section{Methodology}\label{SecIII}
\subsection{Problem Formulation}

Consider a supervised fault-diagnosis dataset
\begin{equation}
\mathcal{D}=\{(\mathbf{x}_i,y_i)\}_{i=1}^{N},
\label{eq:dataset}
\end{equation}
where $\mathbf{x}_i\in\mathbb{R}^{d}$ denotes the feature vector extracted from a monitored system and $y_i\in\{0,1\}$ represents the corresponding fault label for the binary classification task. Throughout this work, binary classification is adopted for all datasets to provide a consistent evaluation of the proposed framework across different application domains. The objective is to learn a nonlinear mapping
\begin{equation}
f:\mathbb{R}^{d}\rightarrow\{0,1\},
\end{equation}
that accurately distinguishes normal and faulty operating conditions while maintaining computational efficiency and robustness against realistic quantum noise. Unlike conventional quantum classifiers that directly infer class labels from quantum measurements, the proposed QSVM-RQNN framework integrates quantum similarity learning with recurrent quantum feature extraction. Specifically, the similarity between an input sample and representative class centroids is first estimated in the quantum feature space, after which the encoded quantum representations are progressively refined through a parameter-shared recurrent quantum circuit. The final classification decision is obtained either from normalized class-conditioned sequence-level similarities in QSVM-RQNN-V1 or from Softmax-normalized marginal qubit measurement scores produced by the recurrent quantum classifier in QSVM-RQNN-V2. Let
\begin{equation}
\hat{y}_i=f(\mathbf{x}_i;\boldsymbol{\Theta}),
\end{equation}
where $\boldsymbol{\Theta}$ denotes the complete set of trainable parameters of the proposed framework. The learning objective is to determine the optimal parameter set that minimizes the empirical classification loss over the training dataset, which can be formulated as
\begin{equation}
\boldsymbol{\Theta}^{*}
=
\arg\min_{\boldsymbol{\Theta}}
\frac{1}{N}
\sum_{i=1}^{N}
\mathcal{L}
\left(
f(\mathbf{x}_i;\boldsymbol{\Theta}),
y_i
\right),
\label{eq:optimization}
\end{equation}
where $\mathcal{L}(\cdot)$ denotes the training objective. For both QSVM-RQNN variants, the training objective is a class-weighted
negative log-likelihood loss. However, the probabilities entering the objective are obtained differently: V1 normalizes two class-conditioned sequence-level similarities, whereas V2 applies Softmax to two marginal qubit measurement scores. The optimization is performed under the architectural constraints imposed by near-term quantum hardware. In particular, the proposed framework employs a fixed three-qubit recurrent quantum circuit with shared parameters across all timesteps, thereby limiting the quantum resource requirements while preserving sufficient representational capacity for fault diagnosis. Accordingly, the optimization problem can be expressed as given in Eq. \eqref{eq:optimization},
\begin{equation}
\begin{aligned}
\textrm{s.t.}\
n_q=3, \
P\leq P_{\max}, \
L\leq L_{\max},
\end{aligned}
\label{eq:constrained_optimization}
\end{equation}
where $n_q$ denotes the number of qubits, $P$ represents the total number of trainable parameters, and $L$ is the depth of the recurrent quantum circuit. This formulation explicitly captures the objective of simultaneously achieving high classification accuracy and low quantum resource consumption, making the proposed QSVM-RQNN framework suitable for implementation on NISQ devices.

\begin{figure}[pos=htbp]
    \centering
    \includegraphics[width=\textwidth]{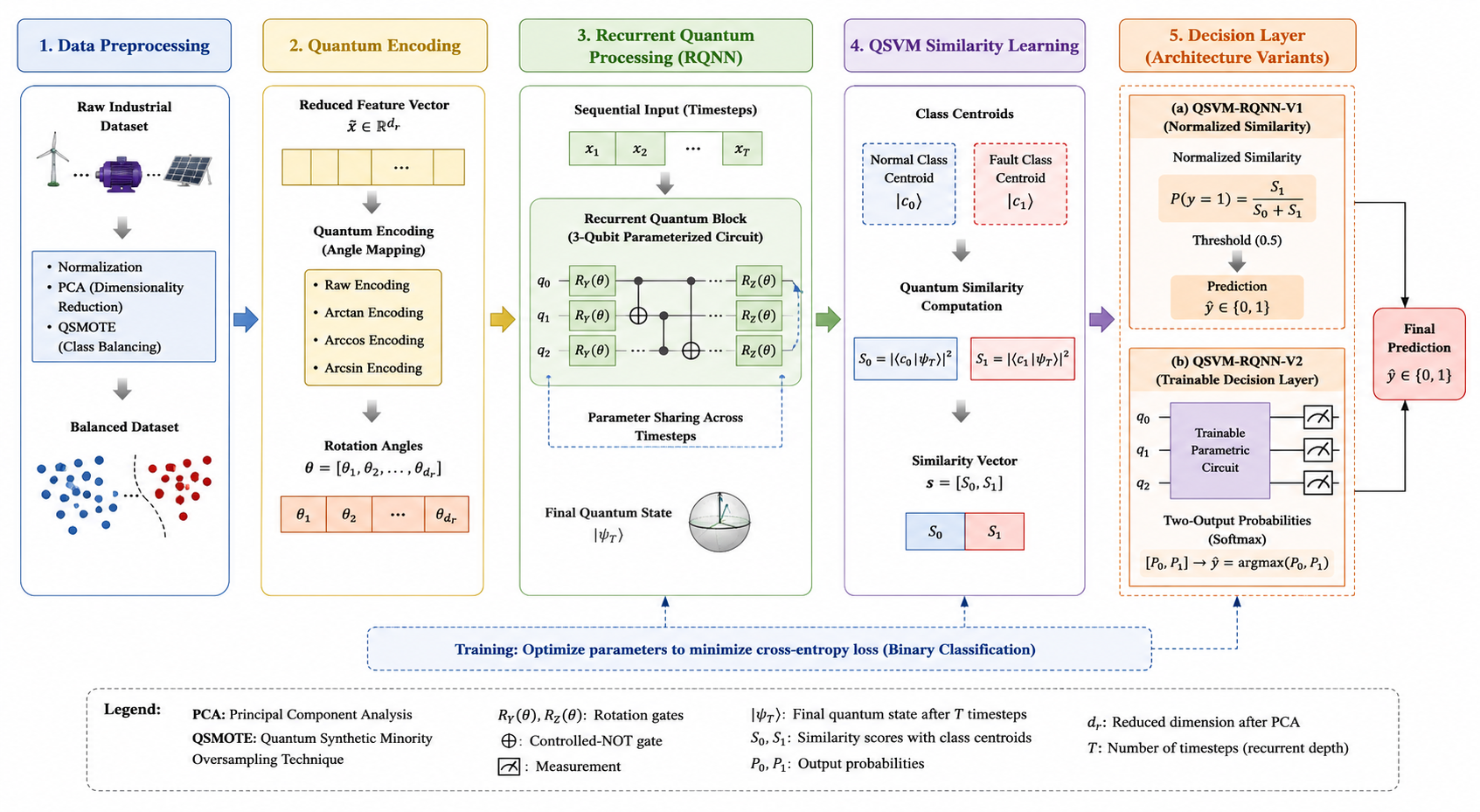}
    \caption{Overall workflow of the proposed QSVM-RQNN framework. The pipeline consists of data preprocessing (normalization, PCA, and QSMOTE), quantum feature encoding, and recurrent quantum representation learning using a compact three-qubit parameterized circuit with shared parameters across timesteps. QSVM-RQNN-V1 performs class-conditioned recurrent similarity learning and classifies using normalized sequence-level quantum similarity scores, whereas QSVM-RQNN-V2 processes timestep-wise quantum similarity features through a recurrent quantum classifier and applies Softmax normalization to the final measurement-derived class scores.}
    \label{fig:framework}
\end{figure}

\subsection{Overall QSVM-RQNN Framework}

The overall workflow of the proposed QSVM-RQNN framework is illustrated in Fig.~\ref{fig:framework}. Given an input feature vector
\(\mathbf{x}\in\mathbb{R}^{d}\), the framework combines classical preprocessing, quantum feature encoding, recurrent quantum representation learning, and quantum similarity learning within a unified hybrid architecture. Although the two proposed variants share the same preprocessing pipeline and recurrent-learning principles, they differ in the manner in which quantum similarity information is incorporated and utilized for the final classification decision. The first stage performs classical preprocessing, including feature normalization, PCA, and class balancing using QSMOTE. The preprocessing operation transforms the original feature vector into a reduced representation

\begin{equation}
\mathbf{z}
=
\mathcal{F}_{\mathrm{Prep}}(\mathbf{x}),
\qquad
\mathbf{z}\in\mathbb{R}^{d_r},
\end{equation}

where \(d_r<d\) denotes the reduced feature dimension. The PCA transformation preserves the dominant information contained in the original data while reducing the dimensionality to enable efficient implementation on a fixed three-qubit quantum circuit. The reduced feature vector is subsequently partitioned into \(T\) sequential timesteps,

\begin{equation}
\mathbf{z}
=
\left[
\mathbf{z}^{(1)},
\mathbf{z}^{(2)},
\ldots,
\mathbf{z}^{(T)}
\right],\label{eq:definez}
\end{equation}

which are processed sequentially by the recurrent quantum architecture. Each timestep segment is transformed into quantum rotation angles using one of the adopted encoding strategies (raw, arctangent, arccosine, or arcsine encoding), depending on the characteristics of the dataset. The encoded quantum information is then processed by the recurrent quantum neural network through shared trainable parameters,

\begin{equation}
|\psi_t\rangle
=
\mathcal{F}_{\mathrm{RQNN}}
\left(
|\psi_{t-1}\rangle,
\mathbf{z}^{(t)};
\boldsymbol{\Theta}
\right),
\qquad
t=1,\ldots,T,
\end{equation}

where \(\boldsymbol{\Theta}\) denotes the shared trainable parameters of the recurrent quantum circuit. Since the same parameters are reused at every timestep, the model captures sequential dependencies without increasing the number of trainable parameters with the sequence length. Representative class centroids are computed from the training data and are denoted by

\begin{equation}
\mathcal{C}
=
\{\mathbf{c}_0,\mathbf{c}_1\},
\end{equation}

where \(\mathbf{c}_0\) and \(\mathbf{c}_1\) correspond to the normal and fault classes, respectively. These representative centroids provide compact quantum prototypes that guide the subsequent similarity-learning stage. Although both QSVM-RQNN variants employ the same preprocessing, timestep partitioning, centroid construction, and recurrent-learning framework, they integrate quantum similarity information differently. In QSVM-RQNN-V1, the complete input sequence is recurrently processed separately with respect to each class-centroid sequence, producing two class-conditioned recurrent quantum states whose final all-zero measurement probabilities are normalized to obtain the class-confidence scores. Consequently, similarity estimation is performed after the complete recurrent evolution. In contrast, QSVM-RQNN-V2 first computes the one-qubit overlap between each timestep segment and the corresponding class-centroid segment, producing a two-dimensional similarity sequence. This similarity sequence is subsequently processed by a single three-qubit recurrent quantum classifier, whose final marginal qubit measurements are transformed into normalized class-confidence scores using a Softmax function. Therefore, QSVM-RQNN-V2 incorporates quantum similarity before recurrent quantum classification, whereas QSVM-RQNN-V1 utilizes similarity after recurrent feature extraction. Consequently, the two architectures differ not only in their final decision mechanism but also in the manner in which quantum similarity information is integrated into the recurrent learning process. Nevertheless, both variants retain the same compact three-qubit architecture, shared recurrent parameterization, and quantum similarity learning principles, thereby providing two complementary trade-offs between parameter efficiency and classification flexibility.

\subsection{Data Representation and Encoding}

Let the preprocessed binary classification dataset be denoted by Eq. \eqref{eq:dataset}, where $\mathbf{x}_i\in\mathbb{R}^{d}$ represents the feature vector of the $i$-th sample and
$y_i\in\{0,1\}$ denotes its corresponding class label. Prior to quantum processing, all features are normalized into the interval $[0,1]$ using min-max normalization,
\begin{equation}
x_{ij}^{\mathrm{norm}}
=
\frac{x_{ij}-x_j^{\min}}
{x_j^{\max}-x_j^{\min}},
\end{equation}
where $x_j^{\min}$ and $x_j^{\max}$ denote the minimum and maximum values of the $j$-th feature across the training set, respectively. The normalized feature matrix is represented as
\begin{equation}
\mathbf{X}
=
\left[
\mathbf{x}_1,
\mathbf{x}_2,
\ldots,
\mathbf{x}_N
\right]^T
\in
\mathbb{R}^{N\times d}.
\end{equation}

Since present-day quantum devices possess limited qubit resources, dimensionality reduction is performed before quantum encoding. PCA is employed to project the normalized feature vectors into a lower-dimensional subspace while preserving the dominant variance of the original data. The reduced feature representation is obtained as
\begin{equation}
\mathbf{z}_i
=
\mathbf{W}^{T}
\mathbf{x}_i,
\end{equation}
where
$\mathbf{W}\in\mathbb{R}^{d\times d_r}$ contains the first $d_r$ principal eigenvectors of the covariance matrix. Throughout this work, the reduced dimension is fixed as $d_r=8$, which provides a compact yet discriminative representation suitable for quantum processing. Instead of encoding the entire reduced feature vector simultaneously, the proposed recurrent architecture partitions each sample into a sequence of smaller feature subsets. Let
\begin{equation}
\mathbf{z}_i
=
\left[
z_{i1},
z_{i2},
\ldots,
z_{id_r}
\right]^T
\end{equation}
denote the PCA-reduced feature vector. It is divided into $T$ consecutive timesteps,
\begin{equation}
\mathbf{z}_i
=
\left[
\mathbf{z}_i^{(1)},
\mathbf{z}_i^{(2)},
\ldots,
\mathbf{z}_i^{(T)}
\right],
\end{equation}
where
$
\mathbf{z}_i^{(t)}
\in
\mathbb{R}^{d_r/T},\ t=1,\ldots,T$. Each timestep is processed sequentially by the recurrent quantum circuit using the same shared trainable parameters across all timesteps. The evolving quantum state enables information from previous timestep segments to influence subsequent recurrent representations. This sequential decomposition enables the model to learn progressively refined quantum representations without increasing the number of physical qubits. To investigate the influence of quantum state preparation, four different encoding strategies are considered. Given a normalized scalar feature $z$, the corresponding quantum rotation angle $\theta$ is computed as

\begin{equation}
\theta=
z,
\qquad
\text{(Raw Encoding)}
\end{equation}

\begin{equation}
\theta=
\arctan(z),
\qquad
\text{(Arctangent Encoding)}
\end{equation}

\begin{equation}
\theta=
\arccos(z),
\qquad
\text{(Arccosine Encoding)}
\end{equation}

\begin{equation}
\theta=
\arcsin(z),
\qquad
\text{(Arcsine Encoding)}.
\end{equation}

The choice of encoding determines the geometric distribution of quantum states on the Bloch sphere and therefore directly influences the discriminative capability of the classifier. Rather than adopting a universal encoding scheme, the best-performing encoding is selected independently for each dataset through empirical evaluation. Each encoded timestep is subsequently embedded into a parameterized three-qubit quantum circuit. Let
\begin{equation}
\boldsymbol{\theta}^{(t)}
=
\left[
\theta_1^{(t)},
\theta_2^{(t)},
\ldots,
\theta_m^{(t)}
\right]
\end{equation}
represent the encoded feature vector associated with timestep $t$, where $m=d_r/T$. Starting from the initial quantum state $|\psi_0\rangle
=
|000\rangle$,
feature-dependent rotation operators are applied to construct the encoded quantum state according to
\begin{equation}
|\psi_t\rangle
=
U_{\mathrm{enc}}
\left(
\boldsymbol{\theta}^{(t)}
\right)
|000\rangle.
\end{equation}

The encoding operator consists of successive single-qubit rotations about the $y$- and $z$-axes,
\begin{equation}
U_{\mathrm{enc}}
=
\prod_{k=1}^{m}
R_Z(\theta_k^{(t)})
R_Y(\theta_k^{(t)}),
\end{equation}
where the rotations are cyclically assigned to the available qubits whenever the number of encoded features exceeds the number of physical qubits. Consequently, every timestep is represented by a compact quantum state while maintaining a fixed three-qubit architecture throughout all experiments. The resulting sequence of encoded quantum states,
\begin{equation}
\left\{
|\psi_1\rangle,
|\psi_2\rangle,
\ldots,
|\psi_T\rangle
\right\},
\end{equation}
forms the input to the proposed recurrent quantum neural network described in the following subsection.

\subsection{Quantum Similarity Learning}

The similarity quantities introduced in this subsection are incorporated differently in the two proposed architectures. In QSVM-RQNN-V1, class-centroid information is included throughout the class-conditioned recurrent evolution, and the final all-zero measurement probabilities serve as sequence-level similarities. In QSVM-RQNN-V2, the overlap with each centroid is evaluated independently at every timestep, and the resulting similarity pairs form the sequential input to the recurrent quantum classifier. Let the binary training dataset consist of two classes,
\begin{equation}
\mathcal{D}
=
\mathcal{D}_0
\cup
\mathcal{D}_1,
\end{equation}
where
\begin{equation}
\mathcal{D}_c
=
\left\{
\mathbf{z}_i
\mid
y_i=c
\right\},
\qquad
c\in\{0,1\}.
\end{equation}

For each class, a representative centroid is computed in the reduced feature space as

\begin{equation}
\mathbf{c}_c
=
\frac{1}{|\mathcal{D}_c|}
\sum_{\mathbf{z}_i\in\mathcal{D}_c}
\mathbf{z}_i,
\qquad
c\in\{0,1\},
\end{equation}

where $|\mathcal{D}_c|$ denotes the number of samples belonging to class $c$. Similar to the input samples, each centroid is partitioned into $T$ timesteps,

\begin{equation}
\mathbf{c}_c
=
\left[
\mathbf{c}_c^{(1)},
\mathbf{c}_c^{(2)},
\ldots,
\mathbf{c}_c^{(T)}
\right].
\label{eq:centroids}
\end{equation}

Consequently, both the input sequence and the corresponding class centroids possess identical temporal structures, enabling timestep-wise similarity evaluation. After quantum encoding, the input sample and each class centroid are represented as

\begin{equation}
|\psi_t\rangle
=
U_{\mathrm{enc}}
(\mathbf{z}^{(t)})
|000\rangle,
\end{equation}

and

\begin{equation}
|\phi_c^{(t)}\rangle
=
U_{\mathrm{enc}}
(\mathbf{c}_c^{(t)})
|000\rangle,
\qquad
c\in\{0,1\},
\end{equation}

respectively. Instead of explicitly computing a quantum kernel matrix, the proposed framework estimates the similarity between the encoded sample and each class centroid through the overlap of their corresponding quantum states. The similarity for class $c$ at timestep $t$ is defined as

\begin{equation}
S_c^{(t)}
=
\left|
\left<
\phi_c^{(t)}
\middle|
\psi_t
\right>
\right|^2,
\end{equation}

which corresponds to the squared fidelity between the two quantum states. Since quantum state fidelity naturally lies within the interval

\begin{equation}
0
\le
S_c^{(t)}
\le
1,
\end{equation}

it provides a normalized measure of geometric similarity in the Hilbert space. In the practical implementation, the overlap is estimated through quantum state evolution. Let

\begin{equation}
U_{\mathrm{sim}}
=
U_{\mathrm{enc}}^\dagger
(\mathbf{c}_c^{(t)})
U_{\mathrm{enc}}
(\mathbf{z}^{(t)}),
\end{equation}

denote the combined quantum transformation. Starting from the initial state $|000\rangle$, the probability of measuring the all-zero computational basis state becomes

\begin{equation}
P_c^{(t)}
=
\left|
\left<
000
\right|
U_{\mathrm{sim}}
\left|
000
\right>
\right|^2,
\end{equation}

which is equivalent to the squared overlap between the encoded sample and the corresponding centroid state,

\begin{equation}
P_c^{(t)}
=
S_c^{(t)}.
\end{equation}

Since similarity is computed with respect to both class centroids, two similarity scores are obtained, $P_0^{(t)},
\ P_1^{(t)}$. To eliminate scale variations and obtain a probabilistic interpretation, the similarity scores are normalized according to

\begin{equation}
\hat{S}_0^{(t)}
=
\frac{P_0^{(t)}}
{P_0^{(t)}+P_1^{(t)}+\varepsilon},
\end{equation}

and

\begin{equation}
\hat{S}_1^{(t)}
=
\frac{P_1^{(t)}}
{P_0^{(t)}+P_1^{(t)}+\varepsilon},
\end{equation}

where $\varepsilon$ is a small positive constant introduced for numerical stability. Unlike conventional QSVMs that rely on constructing and optimizing an entire quantum kernel matrix, the proposed similarity-learning mechanism only evaluates the similarity between an input sample and two representative class centroids. This substantially reduces the computational overhead while preserving the discriminative geometric information required for classification. Furthermore, the normalized similarity scores provide complementary information to the recurrent quantum feature extractor and constitute the primary decision variables in QSVM-RQNN-V1, while also serving as informative quantum representations in QSVM-RQNN-V2.
For notational convenience, the normalized similarity quantities introduced in this subsection are denoted by $\hat{P}_c$ in the architecture-specific descriptions of QSVM-RQNN-V1 and QSVM-RQNN-V2.

\subsection{Recurrent Quantum Neural Network}

Unlike conventional quantum neural networks that encode the entire feature vector into a single parameterized quantum circuit, the proposed QSVM-RQNN framework adopts a recurrent quantum processing strategy in which the input representation is sequentially processed through multiple timesteps using a shared parameterized quantum circuit. This recurrent formulation enables progressive feature refinement while maintaining a fixed number of physical qubits and trainable parameters, making the architecture particularly suitable for near-term quantum hardware. Let the PCA-reduced feature vector of an input sample be represented as given in Eq. \eqref{eq:definez}, 
where each
\(
\mathbf{z}^{(t)}
\in
\mathbb{R}^{m}
\)
corresponds to the feature subset associated with the
$t$-th timestep. Every timestep is independently encoded into a quantum state according to

\begin{equation}
|\psi_t^{(0)}\rangle
=
U_{\mathrm{enc}}
\left(
\mathbf{z}^{(t)}
\right)
|000\rangle,
\end{equation}

where
\(U_{\mathrm{enc}}\)
denotes the quantum encoding operator introduced in the previous subsection. The encoded quantum state is subsequently transformed by a parameterized recurrent quantum circuit,

\begin{equation}
|\psi_t\rangle
=
U_{\mathrm{RQNN}}
(\boldsymbol{\Theta})
|\psi_t^{(0)}\rangle,
\end{equation}

where $\boldsymbol{\Theta}
=
\{\theta_1,\theta_2,\ldots,\theta_P\}$ collectively denotes all trainable circuit parameters. A distinguishing characteristic of the proposed architecture is that the same parameter set
\(
\boldsymbol{\Theta}
\)
is reused across every timestep,

\begin{equation}
U_{\mathrm{RQNN}}^{(1)}
=
U_{\mathrm{RQNN}}^{(2)}
=
\cdots
=
U_{\mathrm{RQNN}}^{(T)}.
\end{equation}

Consequently, the recurrent behavior is achieved through parameter sharing rather than increasing the circuit size or introducing additional trainable variables. This design significantly reduces the overall model complexity while enabling consistent feature extraction throughout the entire input sequence. Each recurrent block consists of alternating parameterized single-qubit rotations followed by entangling operations. For a single recurrent layer, the unitary transformation is expressed as

\begin{equation}
U_{\mathrm{RQNN}}
=
U_{\mathrm{Ent}}
U_{R_Z}
U_{R_Y},
\end{equation}

where

\begin{equation}
U_{R_Y}
=
\prod_{q=1}^{n_q}
R_Y(\theta_q),
\end{equation}

and

\begin{equation}
U_{R_Z}
=
\prod_{q=1}^{n_q}
R_Z(\theta_{n_q+q}),
\end{equation}

with
\(n_q=3\)
denoting the number of qubits. The entangling operation is implemented through a cyclic chain of controlled-NOT gates,

\begin{equation}
U_{\mathrm{Ent}}
=
CX_{1,2}
CX_{2,3}
CX_{3,1},
\end{equation}

which distributes information among all qubits while preserving a compact circuit topology. The complete recurrent transformation is therefore written as

\begin{equation}
U_{\mathrm{RQNN}}
=
\prod_{\ell=1}^{L}
\left(
U_{\mathrm{Ent}}
U_{R_Z}^{(\ell)}
U_{R_Y}^{(\ell)}
\right),
\end{equation}

where
\(L\)
denotes the number of recurrent quantum layers. Accordingly, the final quantum representation of the
$t$-th timestep becomes

\begin{equation}
|\psi_t\rangle
=
\left[
\prod_{\ell=1}^{L}
U_{\mathrm{Ent}}
U_{RZ}^{(\ell)}
U_{RY}^{(\ell)}
\right]
U_{\mathrm{enc}}
\!\left(
\boldsymbol{\xi}^{(t)}
\right)
|\psi_{t-1}\rangle,\ t=1,\ldots,T,
\label{eq:rqnn}
\end{equation}

The quantity $\boldsymbol{\xi}^{(t)}$ denotes the architecture-dependent information supplied to the quantum encoding operator at timestep $t$. In QSVM-RQNN-V1, $\boldsymbol{\xi}^{(t)}$ jointly represents the input timestep and the corresponding class-centroid segment, whereas in QSVM-RQNN-V2, $\boldsymbol{\xi}^{(t)}$ represents the timestep-wise quantum similarity features computed with respect to the two class centroids, as described in Section~\ref{iiig}. The same parameterized recurrent transformation is applied sequentially at every timestep while sharing an identical set of trainable parameters across the complete sequence, $
\{
|\psi_1\rangle,
|\psi_2\rangle,
\ldots,
|\psi_T\rangle
\}$, thereby producing a sequence of recurrent quantum feature representations. Unlike classical recurrent neural networks, where temporal dependencies are propagated through hidden states, the proposed framework employs repeated application of a shared quantum circuit across successive feature partitions. This enables progressive refinement of quantum representations while maintaining constant circuit depth and parameter dimensionality across all timesteps. The number of trainable parameters is therefore determined solely by the number of recurrent layers and qubits,

\begin{equation}
P
=
2Ln_q,
\end{equation}

which yields $P=12$ for the default configuration
\(
L=2
\)
and
\(
n_q=3
\),
consistent with the implemented QSVM-RQNN-V1 architecture. The corresponding quantum representations generated by the recurrent network subsequently serve as the basis for the two decision mechanisms introduced in the following subsections.

\subsection{QSVM-RQNN-V1}

The first proposed architecture, referred to as QSVM-RQNN-V1, combines quantum similarity learning with recurrent quantum feature extraction through a similarity-based decision mechanism. Unlike conventional quantum neural networks that employ an additional trainable output layer, QSVM-RQNN-V1 directly performs classification using normalized quantum similarity scores obtained from the final recurrent quantum representations. Consequently, the architecture retains a compact parameterization while preserving its discriminative capability. Following Eq.~\eqref{eq:centroids}, let

\begin{equation}
\mathbf{c}_0
=
\left[
\mathbf{c}_0^{(1)},
\mathbf{c}_0^{(2)},
\ldots,
\mathbf{c}_0^{(T)}
\right]
\end{equation}

and

\begin{equation}
\mathbf{c}_1
=
\left[
\mathbf{c}_1^{(1)},
\mathbf{c}_1^{(2)},
\ldots,
\mathbf{c}_1^{(T)}
\right]
\end{equation}

represent the class-centroid sequences corresponding to the normal and fault classes, respectively. The centroids are computed from the training samples after preprocessing and are partitioned into the same number of timesteps as the input sequence. For an input sequence given in Eq. \eqref{eq:definez}, the recurrent quantum circuit is evaluated separately with respect to each class-centroid sequence. For class \(c\in\{0,1\}\), the input segment \(\mathbf{z}^{(t)}\) and the corresponding centroid segment \(\mathbf{c}_c^{(t)}\) are encoded at timestep \(t\), followed by the shared recurrent quantum transformation. The recurrent evolution can be expressed as

\begin{equation}
\left|\psi_{c,t}\right\rangle
=
U_{\mathrm{RQNN}}\!\left(\boldsymbol{\Theta}\right)
U_{\mathrm{enc}}\!\left(
\mathbf{z}^{(t)},
\mathbf{c}_c^{(t)}
\right)
\left|\psi_{c,t-1}\right\rangle,\
t=1,2,\ldots,T,
\end{equation}

with the initial state $\left|\psi_{c,0}\right\rangle
=
|000\rangle$. The same trainable parameter vector \(\boldsymbol{\Theta}\) is shared across all timesteps, enabling information to accumulate throughout the recurrent evolution without increasing the number of trainable parameters with \(T\). After all timesteps have been processed, the final class-conditioned recurrent state is

\begin{equation}
\left|\psi_c\right\rangle
=
\left|\psi_{c,T}\right\rangle,
\qquad
c\in\{0,1\}.
\end{equation}

A single quantum measurement is then performed on the final state. The similarity between the input sequence and class \(c\) is obtained from the probability of measuring the all-zero computational basis state,

\begin{equation}
P_c
=
\left|
\left\langle
000
\middle|
\psi_c
\right\rangle
\right|^2,
\qquad
c\in\{0,1\}.
\end{equation}

Therefore, \(P_0\) and \(P_1\) are sequence-level quantum similarity scores obtained after processing the complete recurrent sequence. No intermediate measurement or timestep-wise averaging is performed. To obtain comparable class-confidence scores independent of their absolute magnitudes, the two similarities are normalized as

\begin{equation}
\hat{P}_0
=
\frac{P_0}
{P_0+P_1+\varepsilon},
\end{equation}

and

\begin{equation}
\hat{P}_1
=
\frac{P_1}
{P_0+P_1+\varepsilon},
\end{equation}

where \(\varepsilon\) is a small positive constant introduced for numerical stability. Consequently,

\begin{equation}
\hat{P}_0+\hat{P}_1
=
\frac{P_0+P_1}
{P_0+P_1+\varepsilon}
\approx 1,
\end{equation}

where the negligible deviation from unity is caused only by the numerical stabilizer. The quantities \(\hat{P}_0\) and \(\hat{P}_1\) are interpreted as normalized class-similarity or class-confidence scores rather than posterior probabilities. Unlike conventional QSVM classifiers that construct an explicit quantum kernel matrix followed by classical support-vector optimization, QSVM-RQNN-V1 directly uses the normalized sequence-level quantum similarities for classification. The predicted class is determined using a validation-optimized threshold,

\begin{equation}
\hat{y}
=
\begin{cases}
1, & \hat{P}_1 \geq \tau,\\
0, & \hat{P}_1 < \tau,
\end{cases}
\label{eq:thresholdrule}
\end{equation}

where \(\tau\) denotes the decision threshold. After optimizing the recurrent quantum parameters, \(\tau\) is selected using the validation set to maximize the F1-score. This data-driven thresholding strategy provides an effective trade-off between precision and recall and is particularly useful under class imbalance. It also replaces the fixed maximum-score decision rule, which would be equivalent only when \(\tau=0.5\). To optimize the recurrent quantum parameters, the normalized similarities are incorporated into a weighted negative log-likelihood objective,

\begin{equation}
\mathcal{L}
=
-\frac{1}{N}
\sum_{i=1}^{N}
\left[
w_0(1-y_i)
\log\!\left(\hat{P}_0^{(i)}+\varepsilon\right)
+
w_1y_i
\log\!\left(\hat{P}_1^{(i)}+\varepsilon\right)
\right],
\label{eq:negativelog}
\end{equation}

where \(N\) denotes the total number of training samples and \(N_c\) denotes the number of training samples belonging to class \(c\). The class-balancing weights are defined as

\begin{equation}
w_c
=
\frac{N}
{2N_c},
\qquad
c\in\{0,1\}.
\end{equation}

This weighting strategy reduces the dominance of the majority class and encourages balanced optimization of the normal and fault categories. The additional \(\varepsilon\) inside the logarithm prevents numerical instability when a normalized similarity approaches zero. The trainable parameter vector is written as

\begin{equation}
\boldsymbol{\Theta}
=
\left\{
\theta_1,
\theta_2,
\ldots,
\theta_{12}
\right\},
\end{equation}

and is optimized using the COBYLA optimizer, which is suitable for gradient-free optimization of parameterized quantum circuits. QSVM-RQNN-V1 therefore combines three complementary mechanisms within a unified framework: recurrent quantum representation learning through shared parameterized circuits, quantum similarity estimation using representative class centroids, and direct threshold-based classification without an additional trainable output layer. Algorithm~\ref{alg:qsvm_rqnn_v1} summarizes the corresponding training and inference procedure.

\begin{algorithm}[!t]
\caption{QSVM-RQNN-V1}
\label{alg:qsvm_rqnn_v1}
\begin{algorithmic}[1]
\Require Training dataset $\mathcal{D}_{\mathrm{tr}}$, validation dataset $\mathcal{D}_{\mathrm{val}}$, test dataset $\mathcal{D}_{\mathrm{te}}$, number of timesteps $T$, recurrent depth $L$
\Ensure Predicted test labels $\hat{\mathbf{y}}$

\State Normalize the input features and perform PCA reduction.
\State Partition each reduced feature vector into $T$ timestep segments.
\State Compute the training class centroids $\mathbf{c}_0$ and $\mathbf{c}_1$.
\State Partition each centroid into $T$ timestep segments.
\State Initialize the recurrent parameters $\boldsymbol{\Theta}$.

\While{the COBYLA stopping criterion is not satisfied}
    \For{each training sample $(\mathbf{z}_i,y_i)$}
        \For{$c\in\{0,1\}$}
            \State Initialize $\left|\psi_{c,0}\right\rangle=|000\rangle$.
            \For{$t=1$ to $T$}
                \State Encode $\mathbf{z}_i^{(t)}$ and $\mathbf{c}_c^{(t)}$.
                \State Apply the shared recurrent circuit of depth $L$ using $\boldsymbol{\Theta}$.
            \EndFor
            \State Measure the final all-zero probability
            \[
            P_c^{(i)}
            =
            \left|
            \left\langle
            000
            \middle|
            \psi_{c,T}^{(i)}
            \right\rangle
            \right|^2.
            \]
        \EndFor
        \State Compute
        \[
        \hat{P}_c^{(i)}
        =
        \frac{P_c^{(i)}}
        {P_0^{(i)}+P_1^{(i)}+\varepsilon},
        \qquad
        c\in\{0,1\}.
        \]
    \EndFor
    \State Evaluate the weighted negative log-likelihood loss.
    \State Update $\boldsymbol{\Theta}$ using COBYLA.
\EndWhile

\State Compute validation scores $\hat{P}_1^{(i)}$ using the optimized parameters.
\State Select the threshold $\tau$ that maximizes the validation F1-score.

\For{each test sample $\mathbf{z}_i$}
    \For{$c\in\{0,1\}$}
        \State Process the complete input and centroid sequences through the recurrent circuit.
        \State Measure the final all-zero probability $P_c^{(i)}$.
    \EndFor
    \State Compute the normalized score $\hat{P}_1^{(i)}$.
    \State Predict
    \[
    \hat{y}_i
    =
    \mathbb{I}
    \left(
    \hat{P}_1^{(i)}
    \geq
    \tau
    \right).
    \]
\EndFor

\State \Return Predicted labels $\hat{\mathbf{y}}$.
\end{algorithmic}
\end{algorithm}

\subsection{QSVM-RQNN-V2}\label{iiig}

Although QSVM-RQNN-V1 performs classification directly from normalized
class-conditioned quantum similarities, the second proposed architecture,
referred to as QSVM-RQNN-V2, adopts a two-output recurrent quantum
classification mechanism. In this variant, the similarities between each
timestep of the input sequence and the corresponding class-centroid
segments are first computed using compact one-qubit overlap circuits. These
similarity features are subsequently processed by a three-qubit recurrent
quantum circuit, whose final qubit measurements provide two
class-associated output scores. Let the preprocessed input sample and the two class-centroid sequences be
written as given in Eq. \eqref{eq:definez} and

\begin{equation}
\mathbf{c}_k
=
\left[
\mathbf{c}_k^{(1)},
\mathbf{c}_k^{(2)},
\ldots,
\mathbf{c}_k^{(T)}
\right],
\qquad
k\in\{0,1\},
\end{equation}

respectively. For every timestep \(t\), the input segment
\(\mathbf{z}^{(t)}\) and the centroid segment
\(\mathbf{c}_k^{(t)}\) are amplitude normalized and prepared on a
single-qubit circuit. The timestep-wise QSVM overlap with class \(k\) is
defined as

\begin{equation}
s_k^{(t)}
=
\left|
\left\langle
\phi_k^{(t)}
\middle|
\psi^{(t)}
\right\rangle
\right|^2,
\qquad
k\in\{0,1\},
\end{equation}

where \(\left|\psi^{(t)}\right\rangle\) and
\(\left|\phi_k^{(t)}\right\rangle\) denote the amplitude-encoded input and
centroid states, respectively. In the implementation, this overlap is
obtained as the probability of measuring the state \(\lvert 0\rangle\)
after applying the inverse centroid-state preparation to the encoded input
state. The two similarity values at timestep \(t\),

\begin{equation}
\mathbf{s}^{(t)}
=
\left[
s_0^{(t)},
s_1^{(t)}
\right],
\end{equation}

form the input to the recurrent quantum classifier. They are encoded into
the three-qubit circuit according to

\begin{eqnarray}
U_{\mathrm{sim}}
\left(
\mathbf{s}^{(t)}
\right)
&=&
R_Y^{(0)}
\left(
\pi s_0^{(t)}
\right)
R_Z^{(0)}
\left(
\pi s_0^{(t)}
\right)
R_Y^{(1)}
\left(
\pi s_1^{(t)}
\right)\nonumber\\
&&R_Z^{(1)}
\left(
\pi s_1^{(t)}
\right)
R_Y^{(2)}
\left(
\pi
\left[
s_0^{(t)}-s_1^{(t)}
\right]
\right).
\end{eqnarray}

Following the similarity encoding, each recurrent layer applies
parameterized rotations

\begin{equation}
U_{\mathrm{rot}}^{(\ell)}
\left(
\boldsymbol{\Theta}_{\ell}
\right)
=
\left[
\prod_{q=0}^{2}
R_Y^{(q)}
\left(
\theta_{\ell,q}^{Y}
\right)
R_Z^{(q)}
\left(
\theta_{\ell,q}^{Z}
\right)
\right]
R_X^{(2)}
\left(
\theta_{\ell}^{X}
\right),
\end{equation}

followed by the entangling transformation

\begin{equation}
U_{\mathrm{ent}}
=
\operatorname{CZ}_{2,1}
\operatorname{CZ}_{2,0}
\operatorname{CX}_{1,2}
\operatorname{CX}_{0,2}.
\end{equation}

Accordingly, the recurrent state evolution at timestep \(t\) is expressed
as

\begin{equation}
\left|\psi_t\right\rangle
=
\left[
\prod_{\ell=1}^{L}
U_{\mathrm{ent}}
U_{\mathrm{rot}}^{(\ell)}
\left(
\boldsymbol{\Theta}_{\ell}
\right)
\right]
U_{\mathrm{sim}}
\left(
\mathbf{s}^{(t)}
\right)
\left|\psi_{t-1}\right\rangle,
\end{equation}

with $\left|\psi_0\right\rangle
=
|000\rangle$. The same recurrent parameters are reused at every timestep, allowing the
quantum state to accumulate information from the complete sequence without
increasing the number of trainable parameters with \(T\). After processing
all timesteps, the final recurrent state is

\begin{equation}
\left|\psi_{\mathrm{out}}\right\rangle
=
\left|\psi_T\right\rangle.
\end{equation}

The final state is measured in the computational basis. Two
class-associated quantum scores are then obtained from the marginal
probabilities of qubits \(q_0\) and \(q_1\),

\begin{equation}
o_0
=
\Pr(q_0=1)
=
\sum_{b_2,b_1}
\Pr(b_2b_1 1),
\end{equation}

and

\begin{equation}
o_1
=
\Pr(q_1=1)
=
\sum_{b_2,b_0}
\Pr(b_2 1 b_0).
\end{equation}

Although \(o_0\) and \(o_1\) are individually valid quantum measurement
probabilities, they do not generally form a normalized categorical
distribution because the two marginal measurement events are not mutually
exclusive. Therefore,

\begin{equation}
o_0+o_1
\neq
1
\end{equation}

in general. The two measurement-derived class scores are consequently transformed using the Softmax function,

\begin{equation}
\hat{P}_c
=
\frac{
\exp(o_c)
}{
\exp(o_0)+\exp(o_1)
},
\qquad
c\in\{0,1\},
\end{equation}

such that

\begin{equation}
\hat{P}_0+\hat{P}_1=1.
\end{equation}

The parameters are optimized by minimizing the weighted negative log-likelihood objective as given in Eq. \eqref{eq:negativelog}. The class-balancing weights reduce the influence of class imbalance, while \(\varepsilon\) prevents numerical instability when a predicted probability approaches zero. The complete trainable parameter vector is

\begin{equation}
\boldsymbol{\Theta}
=
\left\{
\boldsymbol{\Theta}_1,
\boldsymbol{\Theta}_2,
\ldots,
\boldsymbol{\Theta}_L
\right\},
\end{equation}

with $\left|
\boldsymbol{\Theta}
\right|
=
7L$. For the implemented configuration \(L=2\), $\left|
\boldsymbol{\Theta}
\right|
=
14$. These parameters are optimized using the COBYLA optimizer. After
optimization, the positive-class probability \(\hat{P}_1\) is evaluated on
the validation set, and the decision threshold \(\tau\) is selected to
maximize the validation F1-score. The final prediction rule is defined in Eq. \eqref{eq:thresholdrule}. Therefore, QSVM-RQNN-V2 differs from QSVM-RQNN-V1 in both its similarity integration and its final decision mechanism. QSVM-RQNN-V1 evaluates two
complete class-conditioned recurrent circuits and directly normalizes
their final all-zero probabilities. In contrast, QSVM-RQNN-V2 first
constructs two timestep-wise QSVM overlap features, jointly processes them
through a single recurrent quantum classifier, and obtains the final class
scores from two qubit marginal measurements. Softmax normalization and
validation-based thresholding are then used to produce the final binary
decision. Algorithm~\ref{alg:qsvm_rqnn_v2} summarizes the implemented
training and inference procedure.

\begin{algorithm}[!t]
\caption{QSVM-RQNN-V2}
\label{alg:qsvm_rqnn_v2}
\begin{algorithmic}[1]
\Require Training dataset $\mathcal{D}_{\mathrm{tr}}$, validation dataset
$\mathcal{D}_{\mathrm{val}}$, test dataset $\mathcal{D}_{\mathrm{te}}$,
number of timesteps $T$, recurrent depth $L$
\Ensure Predicted test labels $\hat{\mathbf{y}}$

\State Normalize the input features and perform PCA reduction.
\State Partition each reduced feature vector into \(T\) timestep segments.
\State Compute the two class centroids using the training split.
\State Partition each centroid into \(T\) timestep segments.

\For{each training, validation, and test sample}
    \For{$t=1$ to $T$}
        \State Compute the one-qubit QSVM overlaps
        \(s_0^{(t)}\) and \(s_1^{(t)}\).
    \EndFor
    \State Form the similarity sequence
    \(\{\left(s_0^{(t)},s_1^{(t)}\right)\}_{t=1}^{T}\).
\EndFor

\State Initialize the recurrent parameters $\boldsymbol{\Theta}$.

\While{the COBYLA stopping criterion is not satisfied}
    \For{each training similarity sequence}
        \State Initialize the three-qubit state as \(\lvert000\rangle\).
        \For{$t=1$ to $T$}
            \State Encode \(s_0^{(t)}\), \(s_1^{(t)}\), and
            \(s_0^{(t)}-s_1^{(t)}\).
            \State Apply the shared recurrent layers using
            \(\boldsymbol{\Theta}\).
        \EndFor
        \State Measure the marginal probabilities
        \(o_0=\Pr(q_0=1)\) and \(o_1=\Pr(q_1=1)\).
        \State Compute
        \[
        \hat{P}_c
        =
        \frac{\exp(o_c)}
        {\exp(o_0)+\exp(o_1)},
        \qquad
        c\in\{0,1\}.
        \]
    \EndFor
    \State Evaluate the weighted negative log-likelihood loss.
    \State Update \(\boldsymbol{\Theta}\) using COBYLA.
\EndWhile

\State Evaluate \(\hat{P}_1\) on the validation set.
\State Select the threshold \(\tau\) that maximizes validation F1-score.

\For{each test sample}
    \State Compute the timestep-wise QSVM overlap sequence.
    \State Process the complete sequence through the trained recurrent
    circuit.
    \State Measure \(o_0\) and \(o_1\) and apply Softmax.
    \State Predict
    \[
    \hat{y}
    =
    \mathbb{I}
    \left(
    \hat{P}_1
    \geq
    \tau
    \right).
    \]
\EndFor

\State \Return Predicted labels $\hat{\mathbf{y}}$.
\end{algorithmic}
\end{algorithm}

\subsection{Computational Complexity}

The computational complexity of the proposed QSVM-RQNN framework is analyzed by considering its principal computational stages, namely classical preprocessing, quantum state encoding, recurrent quantum feature extraction, quantum similarity evaluation, and parameter optimization. Let $N$ denote the number of training samples, $d$ the original feature dimension, $d_r$ the reduced feature dimension after PCA, $T$ the number of recurrent timesteps, $L$ the number of recurrent quantum layers, $n_q$ the number of qubits, and $P$ the total number of trainable parameters. The classical preprocessing stage consists of feature normalization, PCA, and sequence generation. Min-max normalization requires a single traversal of the feature matrix, resulting in

\begin{equation}
\mathcal{O}(Nd).
\end{equation}

The PCA transformation is performed only once before training. Assuming $d \ll N$, its computational cost is dominated by covariance matrix computation and eigenvalue decomposition,

\begin{equation}
\mathcal{O}(Nd^{2}+d^{3}),
\end{equation}

after which each sample is projected into a reduced feature space of dimension $d_r$. Since PCA is an offline preprocessing step, this computational cost is incurred only once. The reduced feature vector is subsequently partitioned into $T$ sequential segments without introducing additional arithmetic operations, resulting in

\begin{equation}
\mathcal{O}(Td_r).
\end{equation}

During quantum state preparation, the reduced feature representation is encoded into a fixed three-qubit quantum circuit using parameterized single-qubit rotations. Since the reduced feature dimension remains fixed throughout all experiments, the encoding complexity is

\begin{equation}
\mathcal{O}(d_r).
\end{equation}

Unlike conventional quantum neural networks that often require increasing numbers of qubits as the input dimension grows, the proposed framework employs a fixed three-qubit architecture. Consequently, the number of physical qubits is independent of the original feature dimension. At each timestep, the recurrent quantum circuit consists of parameterized rotation gates followed by cyclic entangling operations. The computational complexity of a single recurrent block is therefore

\begin{equation}
\mathcal{O}(Ln_q),
\end{equation}

where $L$ denotes the recurrent depth. Processing the complete sequence yields

\begin{equation}
\mathcal{O}(TLn_q).
\end{equation}

The two proposed architectures employ different recurrent parameterizations. QSVM-RQNN-V1 contains two trainable rotation parameters per qubit in each recurrent layer, resulting in

\begin{equation}
P_{\mathrm{V1}}=2Ln_q,
\end{equation}

whereas QSVM-RQNN-V2 introduces one additional trainable rotation within each recurrent layer, leading to

\begin{equation}
P_{\mathrm{V2}}=7L.
\end{equation}

For the implemented configuration with $L=2$ and $n_q=3$, QSVM-RQNN-V1 contains 12 trainable parameters, while QSVM-RQNN-V2 contains 14 trainable parameters. The quantum similarity learning stage requires only two centroid-related similarity evaluations for binary classification at each timestep. Consequently, the similarity computation scales linearly with the sequence length,

\begin{equation}
\mathcal{O}(2T)
=
\mathcal{O}(T),
\end{equation}

which is substantially more efficient than constructing the $N\times N$ quantum kernel matrix required by conventional QSVM approaches. Combining the encoding, recurrent processing, and similarity evaluation stages, the computational complexity of a single forward pass is

\begin{equation}
\mathcal{O}
\left(
d_r
+
TLn_q
+
T
\right).
\end{equation}

Since $d_r$, $T$, and $n_q$ remain fixed throughout all experiments, the forward complexity simplifies to

\begin{equation}
\mathcal{O}(L).
\end{equation}

Although QSVM-RQNN-V1 and QSVM-RQNN-V2 differ in their similarity integration and classification mechanisms, both architectures preserve the same asymptotic forward complexity because they employ identical recurrent depth, fixed three-qubit circuits, and shared parameters across all timesteps. Assuming $K$ optimization iterations over $N$ training samples, the overall training complexity becomes

\begin{equation}
\mathcal{O}
\left(
KNTL
\right),
\end{equation}

which scales linearly with the number of training samples and recurrent layers. The memory complexity is dominated by storing the reduced feature matrix together with the trainable parameters,

\begin{equation}
\mathcal{O}(Nd_r+P).
\end{equation}

Because the proposed framework employs a fixed three-qubit architecture with shared recurrent parameters, both the parameter complexity and memory requirements remain independent of the original feature dimensionality after PCA. Moreover, the centroid-based similarity formulation avoids explicit quantum kernel matrix construction, reducing computational overhead compared with conventional kernel-based QSVMs. Consequently, the proposed QSVM-RQNN framework provides an effective balance between expressive quantum representation learning and computational efficiency, making it suitable for near-term quantum devices with limited resources.

\begin{figure}[pos=htbp]
\centering
\includegraphics[width=\textwidth]{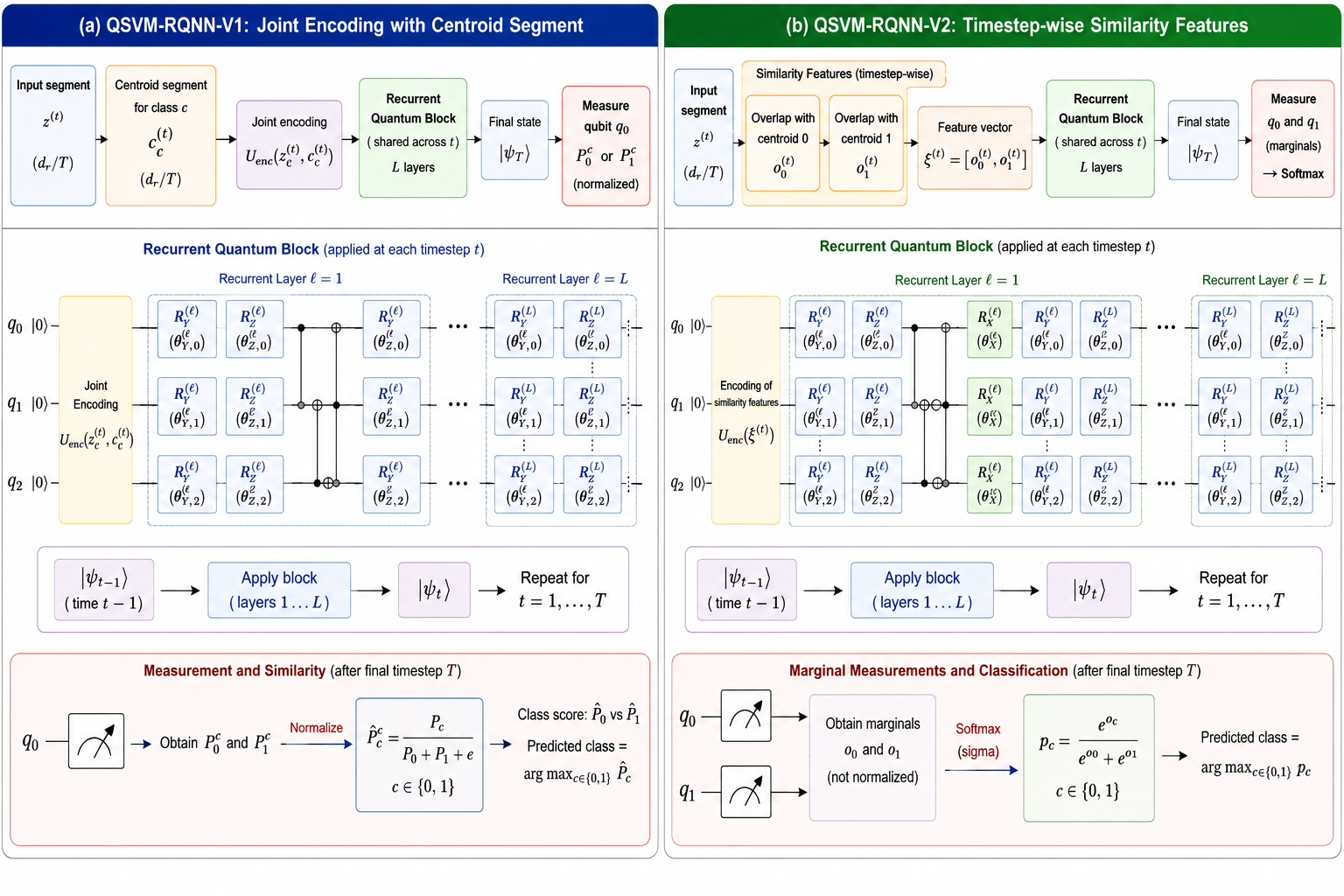}
\caption{Architectural comparison of the proposed QSVM-RQNN variants. (a) \textbf{QSVM-RQNN-V1} performs class-conditioned joint quantum encoding of the input timestep and the corresponding class-centroid segment, followed by recurrent quantum processing using a shared low-qubit recurrent quantum block. After processing the complete sequence, a single final quantum measurement produces sequence-level similarity scores, which are normalized to obtain the class probabilities. (b) \textbf{QSVM-RQNN-V2} first computes timestep-wise quantum similarity features with respect to the two class centroids and encodes these similarity representations into the recurrent quantum circuit. The recurrent block contains an additional trainable $R_X$ rotation in each layer, and the final class probabilities are obtained from marginal qubit measurements followed by Softmax normalization. Both architectures employ the same compact three-qubit recurrent quantum backbone with shared parameters across all timesteps, while differing in the integration of similarity information and the final classification strategy.}
\label{fig:architecture_comparison}
\end{figure}

\subsection{Architectural comparison of V1 and V2}

Figure~\ref{fig:architecture_comparison} highlights the architectural differences between the two proposed QSVM-RQNN variants. Both models share the same compact three-qubit recurrent quantum backbone with parameter sharing across all timesteps, thereby maintaining low parameter complexity and computational efficiency. The primary distinction lies in the representation supplied to the recurrent quantum circuit. QSVM-RQNN-V1 jointly encodes each input timestep together with the corresponding class-centroid segment, enabling the recurrent circuit to learn class-conditioned sequence representations before producing sequence-level similarity scores through a final quantum measurement. In contrast, QSVM-RQNN-V2 first computes timestep-wise quantum similarity features relative to the class centroids and subsequently processes these similarity representations using the recurrent quantum circuit. The final prediction is obtained from marginal qubit measurements followed by Softmax normalization. Consequently, QSVM-RQNN-V1 performs recurrent learning directly on joint input--centroid representations, whereas QSVM-RQNN-V2 performs recurrent learning on similarity representations, providing two complementary mechanisms for integrating quantum similarity learning with recurrent quantum sequence modeling.

\section{Experimental Results}\label{SecIV}

\subsection{Datasets}

The proposed QSVM-RQNN framework is evaluated on four publicly available benchmark datasets covering different fault diagnosis applications, including photovoltaic panel inspection, bearing fault diagnosis, engine fault detection, and industrial equipment monitoring. These datasets provide diverse data modalities, varying class distributions, and multiple fault categories, enabling a comprehensive assessment of the proposed method under different industrial scenarios. A summary of the datasets employed in this work is presented in Table~\ref{tab:datasets}. Although the original benchmark datasets contain multiple operational and fault categories, the experimental evaluation in this work considers binary fault-classification tasks derived from these datasets. For each benchmark, one reference class and one minority fault class are retained, after which the labels are mapped to $\{0,1\}$, where the minority fault category represents the positive class. This formulation enables a consistent evaluation of QSMOTE-based balancing and minority-fault detection across all quantum models.

\begin{table}[pos=htbp]
\centering
\caption{Summary of the benchmark datasets used to evaluate the proposed QSVM-RQNN framework.}
\label{tab:datasets}
\small
\begin{tabular}{lcccc}
\toprule
\textbf{Dataset} &
\textbf{Application Domain} &
\textbf{Samples} &
\textbf{Classes} &
\textbf{Data Type} \\
\midrule

SPID \cite{afroz2023solar}
& Solar panel fault diagnosis
& 708
& 6
& RGB images\\

CWRUBD \cite{cwru_kaggle}
& Bearing fault diagnosis
& After preprocessing
& Multiple
& Time-series\\

EFDD \cite{engine_failure_kaggle}
& Engine fault detection
& 1000
& 4
& Multivariate sensor data\\

IFDD \cite{industrialfault2024}
& Industrial fault diagnosis
& 1000
& 4
& Industrial IoT sensor data\\

\bottomrule
\end{tabular}
\end{table}

\subsection{Preprocessing}

A unified preprocessing pipeline is adopted to ensure that all benchmark datasets are transformed into a consistent representation before quantum learning. Depending on the data modality, dataset-specific preprocessing is first performed, followed by feature normalization and dimensionality reduction to obtain compact feature vectors suitable for quantum encoding. For the SPID image dataset, the images are divided into training and validation subsets using an 80:20 split. Each image is resized to a fixed spatial resolution before deep visual features are extracted using a pretrained ResNet50 backbone with the final classification layer removed. Global average pooling is employed to generate compact image embeddings, after which PCA is applied to retain the most informative components while substantially reducing the feature dimensionality. The extracted features and corresponding labels are then organized into tabular form for subsequent quantum learning. 

For the CWRUBD dataset, vibration measurements are first segmented into fixed-length windows, from which nine standard statistical time-domain descriptors, including maximum, minimum, mean, standard deviation, root mean square (RMS), skewness, kurtosis, crest factor, and form factor, are computed. The categorical fault labels are converted into contiguous numerical class identifiers, producing a structured feature matrix suitable for supervised classification. The EFDD and IFDD datasets consist of multivariate sensor measurements and are processed using a common tabular preprocessing pipeline. Categorical attributes are converted into numerical representations through deterministic label encoding, while timestamp or non-informative attributes are excluded from model training. Numerical features are retained in floating-point format and class labels are standardized into consecutive integer indices to maintain consistency across all experiments. Finally, for every dataset, the extracted feature vectors are standardized and reduced to an eight-dimensional representation using PCA. This common feature space enables all quantum and hybrid quantum models to operate under identical input conditions, ensuring a fair comparison while allowing efficient encoding into the proposed low-qubit QSVM-RQNN architecture.

\subsection{Comparative Models}

The proposed QSVM-RQNN framework is evaluated against several representative quantum and hybrid QML models to provide a comprehensive assessment of its effectiveness. The comparison includes the QSVM, QNN, QCNN, RQNN, and their hybrid variants, namely QSVM-QNN and QSVM-QCNN. These architectures represent different quantum learning paradigms, including similarity-based classification, variational quantum learning, convolutional quantum feature extraction, recurrent quantum processing, and hybrid combinations of these techniques. By comparing QSVM-RQNN with these established quantum models under identical preprocessing, feature representation, and evaluation protocols, the study systematically investigates the benefits of integrating quantum similarity learning with recurrent quantum feature processing for fault classification.

\subsection{Hyperparameters and Evaluation Metrics}\label{Sec4.1}

To ensure a fair comparison, all quantum and hybrid quantum models are evaluated under an identical experimental protocol. For each dataset, the extracted features are first normalized and reduced to eight principal components using PCA before being encoded into quantum states through the corresponding encoding strategy selected for each model. The train-test partition, preprocessing pipeline, and QSMOTE augmentation procedure are kept identical across all experiments to eliminate variations arising from data preparation. Variational quantum models are trained using the same optimization framework with fixed random initialization and \texttt{random\_state=42} to ensure reproducibility, while the best-performing data encoding strategy for each model is selected based on classification performance.

The proposed QSVM-RQNN architecture employs three qubits throughout all experiments and processes the eight-dimensional feature vector as four sequential timesteps with two features per timestep. Two recurrent variants are investigated: QSVM-RQNN-V1, which performs similarity-based decision making using centroid-conditioned overlap probabilities, and QSVM-RQNN-V2, which adopts a two-output softmax decision mechanism. All competing quantum models are evaluated under the same feature representation and experimental settings to enable a consistent architectural comparison. Model performance is assessed using accuracy, precision, recall, and F1-score, which together provide a comprehensive evaluation of classification effectiveness, particularly for imbalanced fault diagnosis tasks. In addition to these standard metrics, architectural efficiency is analyzed using parameter efficiency ($\eta_P$), qubit efficiency ($\eta_Q$), F1-score improvement after QSMOTE ($\Delta F1$), recall improvement ($\Delta$Recall), and normalized false-negative reduction ($\Delta\mathrm{FNR}_{\mathrm{red}}$).

\subsection{Noise Models}

To evaluate the robustness of the proposed QSVM-RQNN framework under realistic quantum hardware imperfections, six standard quantum noise channels are considered: bit-flip (BF), phase-flip (PF), bit-phase-flip (BPF), depolarizing (DP), amplitude damping (AD), and phase damping (PD)~\cite{mohanty2023vehicle}. The first three channels represent Pauli errors arising from imperfect quantum operations, whereas DP models random Pauli errors occurring with equal probability. AD and PD characterize non-unitary decoherence processes corresponding to energy relaxation and phase decoherence, respectively. These noise channels collectively provide a representative approximation of the dominant error mechanisms encountered in current noisy intermediate-scale quantum (NISQ) devices. Each noise channel is represented using its corresponding Kraus operators $\{E_i\}$, and the evolution of a quantum state $\rho$ is expressed as
\begin{equation}
\zeta=\sum_i E_i \rho E_i^\dagger.
\end{equation}

For the Pauli noise channels (BF, PF, and BPF), the Kraus operators are given by
\begin{equation}
E_0=\sqrt{1-\eta}\,I,\qquad
E_1=\sqrt{\eta}\,\sigma,
\end{equation}
where $\sigma\in\{X,Z,Y\}$ corresponds to the BF, PF, and BPF channels, respectively. For the depolarizing channel,
\begin{align}
E_0 &= \sqrt{1-\eta}\,I,\nonumber\\
E_i &= \sqrt{\eta/3}\,\sigma_i,\qquad
\sigma_i\in\{X,Y,Z\}.
\end{align}

The AD and PD channels are modeled using their conventional Kraus representations, where AD captures spontaneous energy relaxation and PD models coherence loss without energy dissipation~\cite{mohanty2023vehicle}. During robustness evaluation, the noise probability is varied from 0 to 0.5, and each channel is applied immediately before measurement to analyze the stability of the proposed QSVM-RQNN architectures under realistic NISQ noise conditions.

\subsection{Results and Analysis}\label{Sec4.3}

\begin{table}[pos=htbp]
\centering
\caption{Performance comparison of quantum and hybrid quantum models on the SPID dataset before and after QSMOTE using the best-performing encoding.}
\small
\begin{tabular}{llcccccccc}
\toprule
\textbf{Model} & \textbf{Encoding} &
\multicolumn{4}{c}{\textbf{Before QSMOTE}} &
\multicolumn{4}{c}{\textbf{After QSMOTE}} \\
\cmidrule(lr){3-6}
\cmidrule(lr){7-10}
&
&
\textbf{Acc.} &
\textbf{Prec.} &
\textbf{Rec.} &
\textbf{F1} &
\textbf{Acc.} &
\textbf{Prec.} &
\textbf{Rec.} &
\textbf{F1} \\
\midrule

QSVM
& arcsin
& {\textbf{0.9070}}
& {\textbf{1.0000}}
& 0.6000
& {\textbf{0.7500}}
& {{0.7846}}
& 0.7368
& {{0.8750}}
& {{0.8000}} \\

QNN
& arctan
& 0.7674
& 0.5000
& {\textbf{0.9000}}
& 0.6429
& 0.7538
& 0.6818
& {{0.9375}}
& 0.7895 \\

QSVM-QNN
& arctan
& 0.6744
& 0.3750
& 0.6000
& 0.4615
& {\textbf{0.8308}}
& {\textbf{0.8000}}
& {{0.8750}}
& {\textbf{0.8358}} \\

QCNN
& raw
& {{0.8605}}
& {{0.7000}}
& {{0.7000}}
& {{0.7000}}
& 0.7231
& 0.6522
& {{0.9375}}
& 0.7692 \\

QSVM-QCNN
& arctan
& 0.7209
& 0.4500
& {\textbf{0.9000}}
& 0.6000
& 0.6923
& 0.6500
& 0.8125
& 0.7222 \\

RQNN-V1
& arcsin
& 0.5581
& 0.2692
& {\textbf{0.9000}}
& 0.4147
& {{0.8000}}
& {{0.7778}}
& 0.8478
& {{0.8116}} \\

RQNN-V2
& arcsin
& {{0.8605}}
& 0.6667
& {{0.8000}}
& {{0.7273}}
& {{0.8000}}
& {{0.7568}}
& {{0.8750}}
& {{0.8116}} \\

QSVM-RQNN-V1
& raw
& {{0.8837}}
& {{0.8571}}
& 0.6000
& {{0.7059}}
& 0.7231
& 0.6591
& {\textbf{0.9063}}
& 0.7632 \\

QSVM-RQNN-V2
& arccos
& 0.7442
& 0.4706
& {{0.8000}}
& 0.5926
& 0.6154
& 0.5818
& {\textbf{1.0000}}
& 0.7356 \\

\bottomrule
\end{tabular}
\label{tab:spid_quantum_models}
\end{table}

\begin{table}[pos=htbp]
\centering
\caption{Performance comparison of quantum and hybrid quantum models on the CWRUBD before and after QSMOTE using the best-performing encoding.}
\small
\begin{tabular}{llcccccccc}
\toprule
\textbf{Model} & \textbf{Encoding} &
\multicolumn{4}{c}{\textbf{Before QSMOTE}} &
\multicolumn{4}{c}{\textbf{After QSMOTE}} \\
\cmidrule(lr){3-6}
\cmidrule(lr){7-10}
&
&
\textbf{Acc.} &
\textbf{Prec.} &
\textbf{Rec.} &
\textbf{F1} &
\textbf{Acc.} &
\textbf{Prec.} &
\textbf{Rec.} &
\textbf{F1} \\
\midrule

QSVM
& arccos
& 0.9478 & 0.6571 & \textbf{1.0000} & 0.7931
& 0.8696 & 0.9048 & 0.8261 & 0.8636 \\

QNN
& arccos
& 0.9310 & 0.7857 & 0.9167 & 0.8462
& 0.8478 & 0.9211 & 0.7609 & 0.8333 \\

QSVM-QNN
& arccos
& 0.8966 & 0.6667 & \textbf{1.0000} & 0.8000
& 0.8696 & 0.8864 & 0.8478 & 0.8667 \\

QCNN
& arccos
& \textbf{0.9655} & \textbf{0.9167} & 0.9167 & \textbf{0.9167}
& 0.8587 & 0.9024 & 0.8043 & 0.8506 \\

QSVM-QCNN
& raw
& \textbf{0.9655} & \textbf{0.9167} & 0.9167 & \textbf{0.9167}
& 0.8261 & 0.7500 & {{0.9783}} & 0.8491 \\

RQNN-V1
& raw
& 0.9483 & 0.9091 & 0.8333 & 0.8696
& 0.9022 & \textbf{0.9512} & 0.8478 & 0.8966 \\

RQNN-V2
& arccos
& 0.9000 & 0.5000 & 0.8478 & 0.6290
& 0.8043 & 0.9118 & 0.6739 & 0.7750 \\

QSVM-RQNN-V1
& arctan
& 0.9138 & 0.7692 & 0.8333 & 0.8000
& {\textbf{0.9130}} & {0.8653} & {\textbf{0.9782}} & {\textbf{0.9183}} \\

QSVM-RQNN-V2
& arccos
& 0.9310 & 0.9000 & 0.7500 & 0.8182
& 0.8587 & 0.8511 & {{0.8696}} & 0.8602 \\

\bottomrule
\end{tabular}
\label{tab:cwru_quantum_models}
\end{table}

\begin{table}[pos=htbp]
\centering
\caption{Performance comparison of quantum and hybrid quantum models on the EFDD dataset before and after QSMOTE using the best-performing encoding.}
\small
\begin{tabular}{llcccccccc}
\toprule
\textbf{Model} & \textbf{Encoding} &
\multicolumn{4}{c}{\textbf{Before QSMOTE}} &
\multicolumn{4}{c}{\textbf{After QSMOTE}} \\
\cmidrule(lr){3-6}
\cmidrule(lr){7-10}
&
&
\textbf{Acc.} &
\textbf{Prec.} &
\textbf{Rec.} &
\textbf{F1} &
\textbf{Acc.} &
\textbf{Prec.} &
\textbf{Rec.} &
\textbf{F1} \\
\midrule

QSVM
& arcsin
& 0.1613 & 0.1613 & 1.0000 & 0.2778
& 0.5000 & 0.5000 & \textbf{1.0000} & 0.6667 \\

QNN
& arccos
& 0.5968 & 0.2222 & 0.6000 & 0.3243
& 0.4904 & 0.4951 & 0.9808 & 0.6581 \\

QSVM-QNN
& raw
& 0.2419 & 0.1754 & 1.0000 & 0.2985
& 0.5577 & 0.5319 & 0.9615 & 0.6849 \\

QCNN
& arccos
& 0.6129 & 0.2308 & 0.6000 & 0.3333
& 0.5000 & 0.5000 & 0.9808 & 0.6623 \\

QSVM-QCNN
& arctan
& 0.6290 & 0.1176 & 0.2000 & 0.1481
& 0.5288 & 0.5149 & \textbf{1.0000} & 0.6797 \\

RQNN-V1
& arctan
& \textbf{0.8226} & 0.0000 & 0.0000 & 0.0000
& 0.5769 & 0.5426 & 0.9808 & 0.6986 \\

RQNN-V2
& arcsin
& 0.8065 & 0.0000 & 0.0000 & 0.0000
& 0.5577 & 0.5349 & 0.8846 & 0.6667 \\

QSVM-RQNN-V1
& arcsin
& 0.8065 & \textbf{0.3750} & \textbf{0.3000} & \textbf{0.3333}
& {\textbf{0.5865}} & {\textbf{0.5495}} & {{0.9615}} & {\textbf{0.6993}} \\

QSVM-RQNN-V2
& arccos
& 0.7419 & 0.2000 & 0.2000 & 0.2000
& 0.5288 & 0.5152 & {0.9808} & {0.6755} \\

\bottomrule
\end{tabular}
\label{tab:efdd_quantum_models}
\end{table}

\begin{table}[pos=htbp]
\centering
\caption{Performance comparison of quantum and hybrid quantum models on the IFDD dataset before and after QSMOTE using the best-performing encoding.}
\small
\begin{tabular}{llcccccccc}
\toprule
\textbf{Model} & \textbf{Encoding} &
\multicolumn{4}{c}{\textbf{Before QSMOTE}} &
\multicolumn{4}{c}{\textbf{After QSMOTE}} \\
\cmidrule(lr){3-6}
\cmidrule(lr){7-10}
&
&
\textbf{Acc.} &
\textbf{Prec.} &
\textbf{Rec.} &
\textbf{F1} &
\textbf{Acc.} &
\textbf{Prec.} &
\textbf{Rec.} &
\textbf{F1} \\
\midrule

QSVM
& arcsin
& 0.1104 & 0.1104 & 1.0000 & 0.1989
& 0.5000 & 0.5000 & 1.0000 & 0.6667 \\

QNN
& arctan
& 0.4908 & 0.0988 & 0.4444 & 0.1616
& 0.5172 & 0.5088 & 0.9931 & 0.6729 \\

QSVM-QNN
& arcsin
& 0.1166 & 0.1063 & 0.9444 & 0.1910
& 0.5069 & 0.5035 & 1.0000 & 0.6697 \\

QCNN
& arcsin
& 0.7055 & 0.1053 & 0.2222 & 0.1429
& 0.5000 & 0.5000 & 1.0000 & 0.6667 \\

QSVM-QCNN
& arccos
& 0.1779 & 0.1184 & 1.0000 & 0.2118
& 0.5034 & 0.5017 & 1.0000 & 0.6682 \\

RQNN-V1
& raw
& 0.7117 & 0.0857 & 0.1667 & 0.1132
& 0.5103 & 0.5052 & 1.0000 & 0.6713 \\

RQNN-V2
& raw
& {0.6994} & 0.1220 & 0.2778 & 0.1695
& \textbf{0.5379} & \textbf{0.5197} & {1.0000} & \textbf{0.6840} \\

QSVM-RQNN-V1
& arctan
& \textbf{0.7423} & \textbf{0.1667} & 0.3333 & \textbf{0.2222}
& {{0.5241}} & {{0.5124}} & {{1.0000}} & {{0.6776}} \\

QSVM-RQNN-V2
& arcsin
& 0.5337 & 0.1081 & \textbf{0.4444} & 0.1739
& {{0.5138}} & {{0.5070}} & {{1.0000}} & {{0.6729}} \\

\bottomrule
\end{tabular}
\label{tab:ifdd_quantum_models}
\end{table}

The results presented in Table~\ref{tab:spid_quantum_models} demonstrate that QSMOTE substantially improves the performance of most quantum and hybrid quantum models on the SPID dataset by enhancing minority-class representation. While the QSVM model achieved the highest pre-QSMOTE accuracy (0.9070), the best post-QSMOTE performance was obtained by the QSVM-QNN model using arctan encoding, achieving an accuracy of 0.8308 and an F1-score of 0.8358. Recurrent architectures such as RQNN-V1 and RQNN-V2 also benefited significantly from QSMOTE, both achieving an F1-score of 0.8116. From an encoding perspective, arctan encoding was particularly effective for hybrid architectures, whereas arcsin encoding consistently favored recurrent quantum models.

As shown in Table~\ref{tab:cwru_quantum_models}, the CWRUBD exhibited strong baseline performance even before QSMOTE, indicating that the dataset is relatively separable despite class imbalance. The highest pre-QSMOTE accuracy of 0.9655 was achieved by both QCNN and QSVM-QCNN models. After QSMOTE, recurrent architectures become dominant, with QSVM-RQNN-V1 using arctan encoding achieving the best overall performance, including an accuracy of 0.9130, precision of 0.8653, recall of 0.9782, and F1-score of 0.9183. Furthermore, arccos encoding emerged as the preferred encoding for most non-recurrent models, whereas recurrent architectures favored raw and arcsin encodings, highlighting their ability to exploit richer sequential feature representations.

The results reported in Table~\ref{tab:efdd_quantum_models} indicate that the EFDD dataset presents a considerably more challenging classification scenario, where several models exhibited severe class-imbalance effects before QSMOTE. In particular, RQNN-V1 and RQNN-V2 produced relatively high accuracies but zero F1-scores, indicating a complete bias toward majority-class predictions. Following QSMOTE, substantial improvements were observed across all architectures. The best performance was achieved by QSVM-RQNN-V1 with arcsin encoding, yielding an F1-score of 0.6993, closely followed by RQNN-V1 with arctan encoding. These results demonstrate that recurrent quantum architectures derive the greatest benefit from QSMOTE-generated samples when dealing with highly imbalanced industrial fault datasets.

Finally, Table~\ref{tab:ifdd_quantum_models} summarizes the results obtained on the IFDD dataset, which appears to be the most challenging among all datasets considered. Although all models exhibited relatively low performance levels before QSMOTE, significant improvements in recall and F1-score were observed after resampling. The best overall performance was achieved by RQNN-V2 with raw encoding, attaining an accuracy of 0.5379 and an F1-score of 0.6840, while QSVM-RQNN-V1 with arctan encoding achieved the second-highest F1-score of 0.6776. The results across Tables~\ref{tab:spid_quantum_models}--\ref{tab:ifdd_quantum_models} demonstrate that QSMOTE consistently enhances minority-class detection, recurrent quantum architectures become increasingly advantageous as dataset complexity increases, and no single encoding strategy universally dominates, with arctan, arccos, arcsin, and raw encodings each proving optimal under different data characteristics and model architectures.

The results presented in Tables~\ref{tab:spid_quantum_models}-\ref{tab:ifdd_quantum_models} demonstrate that QSMOTE consistently improves the classification performance of all quantum and hybrid quantum models, particularly in terms of recall and F1-score, highlighting its effectiveness in mitigating class imbalance. Across the four datasets, the proposed QSVM-RQNN architectures either achieve the highest values for one or more evaluation metrics or remain among the top-performing models. On the CWRUBD and EFDD datasets, QSVM-RQNN-V1 attains the best post-QSMOTE performance, achieving the highest F1-scores of 0.9183 and 0.6993, respectively, along with leading accuracy, precision, and recall values. For the IFDD dataset, QSVM-RQNN-V1 exhibits the strongest pre-QSMOTE performance, achieving the highest accuracy, precision, and F1-score among all evaluated models, demonstrating superior robustness even under severe class imbalance. On the SPID dataset, although QSVM-QNN achieves the highest post-QSMOTE F1-score, QSVM-RQNN-V2 attains the highest recall of 1.0000, indicating perfect fault detection. These findings show that the proposed QSVM-RQNN framework consistently ranks among the top-performing models across all datasets, validating the effectiveness of integrating QSVM-based similarity learning with recurrent quantum feature processing for fault classification.

The encoding analysis reveals several interesting trends across the evaluated quantum and hybrid quantum models. Although no single encoding consistently dominates all datasets and architectures, clear patterns emerge from the best-performing configurations reported in Tables~\ref{tab:spid_quantum_models}-\ref{tab:ifdd_quantum_models}. For the non-recurrent architectures, including QSVM, QNN, QSVM-QNN, QCNN, and QSVM-QCNN, inverse trigonometric encodings frequently provide the best results. In particular, \textit{arccos} emerges as the most frequently selected encoding on the CWRUBD dataset, producing the best-performing configurations for QSVM, QNN, QSVM-QNN, and QCNN, while \textit{arcsin} performs strongly for QSVM on the SPID, EFDD, and IFDD datasets. These observations suggest that inverse trigonometric transformations provide an effective nonlinear mapping of the feature space for similarity-based and feed-forward quantum learning architectures.

A different trend is observed for the recurrent architectures. The RQNN and QSVM-RQNN models do not rely on a single dominant encoding and instead achieve their best performance using a mixture of \textit{raw}, \textit{arcsin}, \textit{arccos}, and \textit{arctan} encodings depending on the dataset characteristics. For the proposed QSVM-RQNN framework, QSVM-RQNN-V1 achieves its best results using \textit{raw} encoding on SPID, \textit{arctan} encoding on CWRUBD and IFDD, and \textit{arcsin} encoding on EFDD, whereas QSVM-RQNN-V2 favors \textit{arccos} encoding on SPID, CWRUBD, and EFDD, and \textit{arcsin} encoding on IFDD. Notably, \textit{arctan} and \textit{arccos} appear most frequently among the best-performing QSVM-RQNN configurations, suggesting that these encodings are particularly well suited to the integration of QSVM-based similarity learning and recurrent quantum processing. The results indicate that recurrent quantum architectures benefit from greater encoding flexibility than conventional QNN and QCNN models, allowing QSVM-RQNN to effectively adapt its feature representation to diverse condition-monitoring and fault-classification datasets.

\begin{table}[pos=htbp]
\centering
\caption{Architectural comparison of quantum and hybrid quantum models.}
\resizebox{\textwidth}{!}{
\begin{tabular}{lcccccccc}
\toprule
\textbf{Model}
&
\textbf{Qubits}
&
\textbf{PCA Features}
&
\textbf{Timesteps}
&
\textbf{Trainable Parameters}
&
\textbf{Entangling Gates}
&
\textbf{Recurrent}
&
\textbf{QSVM Similarity}
&
\textbf{Decision Strategy}
\\
\midrule

QSVM
& 3 & 8 & -- & 0 & 0 & No & Yes & Similarity-based \\

QNN
& 3 & 8 & -- & 12 & 6 & No & No & Single-output \\

QSVM-QNN
& 3 & 8 & -- & 12 & 6 & No & Yes & Single-output \\

QCNN
& 3 & 8 & -- & 12 & 6 & No & No & Single-output \\

QSVM-QCNN
& 3 & 8 & -- & 12 & 6 & No & Yes & Single-output \\

RQNN-V1
& 3 & 8 & 4 & 12 & 24 & Yes & No & Similarity-based \\

RQNN-V2
& 3 & 8 & 4 & 14 & 32 & Yes & No & Two-output Softmax \\

QSVM-RQNN-V1
& 3 & 8 & 4 & 12 & 24 & Yes & Yes & Similarity-based \\

QSVM-RQNN-V2
& 3 & 8 & 4 & 14 & 32 & Yes & Yes & Two-output Softmax \\

\bottomrule
\end{tabular}
}
\label{tab:architecture_analysis}
\end{table}

\begin{table}[pos=htbp]
\centering
\caption{QSMOTE-induced improvement and resource-efficiency comparison of quantum and hybrid quantum models. Here, $\Delta FNR_{\mathrm{red}}$ denotes normalized false-negative reduction.}
\resizebox{\textwidth}{!}{
\begin{tabular}{llccccc}
\toprule
\textbf{Dataset} & \textbf{Model} &
$\boldsymbol{\Delta F1}$ &
$\boldsymbol{\Delta Recall}$ &
$\boldsymbol{\eta_P=F1_{After}/N_{params}}$ &
$\boldsymbol{\eta_Q=F1_{After}/N_q}$ &
$\boldsymbol{\Delta FNR_{\mathrm{red}}}$ \\
\midrule

SPID & QSVM & 0.0500 & 0.2750 & -- & 0.2667 & 0.2750 \\
 & QNN & 0.1466 & 0.0375 & 0.0658 & 0.2632 & 0.0375 \\
 & QSVM-QNN & 0.3743 & 0.2750 & \textbf{0.0697} & \textbf{0.2786} & 0.2750 \\
 & QCNN & 0.0692 & 0.2375 & 0.0641 & 0.2564 & 0.2375 \\
 & QSVM-QCNN & 0.1222 & -0.0875 & 0.0602 & 0.2407 & -0.0875 \\
 & RQNN-V1 & \textbf{0.3969} & -0.0522 & 0.0676 & 0.2705 & -0.0522 \\
 & RQNN-V2 & 0.0843 & 0.0750 & 0.0580 & 0.2705 & 0.0750 \\ 
 & QSVM-RQNN-V1 & 0.0573 & \textbf{0.3063} & 0.0636 & 0.2544 & \textbf{0.3063} \\
 & QSVM-RQNN-V2 & 0.1430 & 0.2000 & 0.0525 & 0.2452 & 0.2000 \\

\midrule

CWRUBD & QSVM & 0.0705 & -0.1739 & -- & 0.2879 & -0.1739 \\
 & QNN & -0.0129 & -0.1558 & 0.0694 & 0.2778 & -0.1558 \\
 & QSVM-QNN & 0.0667 & -0.1522 & 0.0722 & 0.2889 & -0.1522 \\
 & QCNN & -0.0661 & -0.1124 & 0.0709 & 0.2835 & -0.1124 \\
 & QSVM-QCNN & -0.0676 & 0.0616 & 0.0708 & 0.2830 & 0.0616 \\
 & RQNN-V1 & 0.0270 & 0.0145 & 0.0747 & 0.2989 & 0.0145 \\
 & RQNN-V2 & \textbf{0.1460} & -0.1739 & 0.0554 & 0.2583 & -0.1739 \\
 & QSVM-RQNN-V1 & 0.1183 & \textbf{0.1449} & \textbf{0.0765} & \textbf{0.3061} & \textbf{0.1449} \\
 & QSVM-RQNN-V2 & 0.0420 & 0.1196 & 0.0614 & 0.2867 & 0.1196 \\

\midrule

EFDD & QSVM & 0.3889 & 0.0000 & -- & 0.2222 & 0.0000 \\
 & QNN & 0.3338 & 0.3808 & 0.0548 & 0.2194 & 0.3808 \\
 & QSVM-QNN & 0.3864 & -0.0385 & 0.0571 & 0.2283 & -0.0385 \\
 & QCNN & 0.3290 & 0.3808 & 0.0552 & 0.2208 & 0.3808 \\
 & QSVM-QCNN & 0.5316 & 0.8000 & 0.0566 & 0.2266 & 0.8000 \\
 & RQNN-V1 & \textbf{0.6986} & \textbf{0.9808} & 0.0582 & 0.2329 & \textbf{0.9808} \\
 & RQNN-V2 & 0.6667 & 0.8846 & 0.0476 & 0.2222 & 0.8846 \\
 & QSVM-RQNN-V1 & 0.3660 & 0.6615 & \textbf{0.0583} & \textbf{0.2331} & 0.6615 \\
 & QSVM-RQNN-V2 & 0.4755 & 0.7808 & 0.0483 & 0.2252 & 0.7808 \\

\midrule

IFDD & QSVM & 0.4678 & 0.0000 & -- & 0.2222 & 0.0000 \\
 & QNN & 0.5113 & 0.5487 & 0.0561 & 0.2243 & 0.5487 \\
 & QSVM-QNN & 0.4787 & 0.0556 & 0.0558 & 0.2232 & 0.0556 \\
 & QCNN & 0.5238 & 0.7778 & 0.0556 & 0.2222 & 0.7778 \\
 & QSVM-QCNN & 0.4564 & 0.0000 & 0.0557 & 0.2227 & 0.0000 \\
 & RQNN-V1 & \textbf{0.5581} & \textbf{0.8333} & 0.0559 & 0.2238 & \textbf{0.8333} \\
 & RQNN-V2 & 0.5145 & 0.7222 & 0.0489 & \textbf{0.2280} & 0.7222 \\
 & QSVM-RQNN-V1 & 0.4554 & 0.6667 & \textbf{0.0565} & 0.2259 & 0.6667 \\
 & QSVM-RQNN-V2 & 0.4990 & 0.5556 & 0.0481 & 0.2243 & 0.5556 \\

\bottomrule
\end{tabular}
}
\label{tab:qsmote_gain_efficiency}
\end{table}

\subsection{Architectural Analysis of the Proposed QSVM-RQNN}
The architectural and efficiency analyses presented in Tables~\ref{tab:architecture_analysis} and~\ref{tab:qsmote_gain_efficiency} further highlight the advantages of the proposed QSVM-RQNN framework. As shown in Table~\ref{tab:architecture_analysis}, QSVM-RQNN is the only architecture that simultaneously combines QSVM-based similarity learning and recurrent quantum feature processing while maintaining a low-qubit implementation with only 12-14 trainable parameters. Despite this compact architecture, QSVM-RQNN consistently demonstrates strong performance-efficiency characteristics across multiple fault-diagnosis datasets. In particular, QSVM-RQNN-V1 achieves the highest parameter efficiency ($\eta_P$) on the CWRUBD, EFDD, and IFDD datasets, demonstrating that it extracts more predictive performance per trainable parameter than competing quantum models. Similarly, it achieves the highest qubit efficiency ($\eta_Q$) on the CWRUBD and EFDD datasets, indicating superior utilization of limited quantum resources. Furthermore, QSVM-RQNN-V1 attains the largest recall improvement and false-negative reduction on the SPID and CWRUBD datasets, which is especially important for fault diagnosis applications where missed fault detections can be significantly more costly than false alarms. These results suggest that the integration of QSVM-based similarity estimation and recurrent quantum processing is particularly effective for complex and highly imbalanced fault-classification tasks.

The proposed QSVM-RQNN architecture provides a unified mechanism that combines the margin-sensitive similarity estimation of QSVM with the sequential representation-learning capability of RQNN. Unlike standalone QSVM, which primarily relies on centroid- or kernel-based similarity, the proposed model processes the encoded feature vector as a short temporal sequence after PCA reduction. This enables the circuit to learn feature interactions across multiple timesteps rather than treating the full feature vector as a static input. Compared with QNN and QCNN, the recurrent structure repeatedly applies trainable quantum transformations across feature segments, allowing information propagation through the circuit and improving sensitivity to minority fault patterns after QSMOTE.

The advantage of QSVM-RQNN becomes more visible on complex and highly imbalanced datasets. As shown in Table~\ref{tab:cwru_quantum_models}, QSVM-RQNN-V1 achieves the best post-QSMOTE result on CWRUBD, with an accuracy of 0.9130, recall of 0.9782, and F1-score of 0.9183. Similarly, Table~\ref{tab:efdd_quantum_models} shows that QSVM-RQNN-V1 obtains the best EFDD performance, achieving an F1-score of 0.6993. On IFDD, although RQNN-V2 achieves the highest F1-score, QSVM-RQNN-V1 remains the second-best method, demonstrating competitive robustness on the most challenging dataset. These results indicate that the integrated QSVM-RQNN model is especially useful when the classification task requires both similarity-aware decision boundaries and recurrent feature processing.

The comparison between QSVM-RQNN-V1 and QSVM-RQNN-V2 further highlights the importance of the decision mechanism. QSVM-RQNN-V1 uses centroid-conditioned similarity probabilities and compares normalized class-wise overlap scores, whereas QSVM-RQNN-V2 transforms QSVM similarities into two output scores measured from separate qubits and applies a softmax-based decision rule. Therefore, the difference between the two variants is not merely the encoding strategy, but the final class-score construction. V1 behaves as a similarity-driven recurrent classifier, while V2 behaves as a two-output recurrent quantum classifier. This distinction helps explain why V1 achieves superior overall F1-scores on CWRUBD and EFDD, whereas V2 exhibits stronger recall-oriented behavior on SPID and IFDD.

From a computational perspective, the proposed QSVM-RQNN has moderate quantum resource requirements. For an input reduced to $d=8$ PCA features and reshaped into $T=4$ timesteps with two features per timestep, the circuit uses three qubits and a fixed number of recurrent layers. If $L$ denotes the number of recurrent layers, each timestep requires data-encoding rotations, centroid-conditioned operations, trainable rotations, and entangling gates. Therefore, the overall circuit depth scales approximately as $\mathcal{O}(T L)$, while the number of trainable parameters scales as $\mathcal{O}(L n_q)$, where $n_q$ is the number of qubits. This scaling is more efficient than increasing the number of qubits with the full feature dimension and makes the model suitable for near-term quantum simulation and small-scale NISQ implementation. QSVM-RQNN provides a balanced architecture that combines similarity-driven learning, recurrent quantum feature processing, and low-qubit execution, yielding a favorable trade-off between predictive performance, resource efficiency, and architectural expressiveness compared with existing QSVM-, QNN-, QCNN-, and RQNN-based approaches.

\subsection{Robustness Analysis of QSVM-RQNN}

Among the two proposed variants, QSVM-RQNN-V1 demonstrates stronger overall classification performance, achieving higher accuracy, precision, and F1-score across all evaluated datasets while maintaining competitive recall. QSVM-RQNN-V2, on the other hand, provides a complementary two-output decision mechanism and shows strong recall-oriented behavior in several cases. Therefore, to provide a complete robustness assessment of the proposed QSVM-RQNN framework, both QSVM-RQNN-V1 and QSVM-RQNN-V2 are evaluated under different quantum noise channels after QSMOTE. The robustness analysis is restricted to the proposed QSVM-RQNN variants, while noise robustness studies of standard QSVM, QNN, QCNN, and related baseline architectures can be referred to from existing quantum machine learning literature. It should be noted that the QSMOTE procedure is stochastic and generates different synthetic samples across independent executions. Therefore, the balanced datasets used for the robustness analysis and the primary classification experiments are generated independently. As a result, the F1-score at zero noise probability (\(\eta=0\)) in Figures may exhibit slight differences from the post-QSMOTE values reported in Tables. These variations are expected and reflect the inherent randomness of the oversampling process rather than any effect of the injected quantum noise.

\begin{figure}[pos=htbp]
\centering

\begin{subfigure}{0.48\textwidth}
    \centering
    \includegraphics[width=\linewidth]{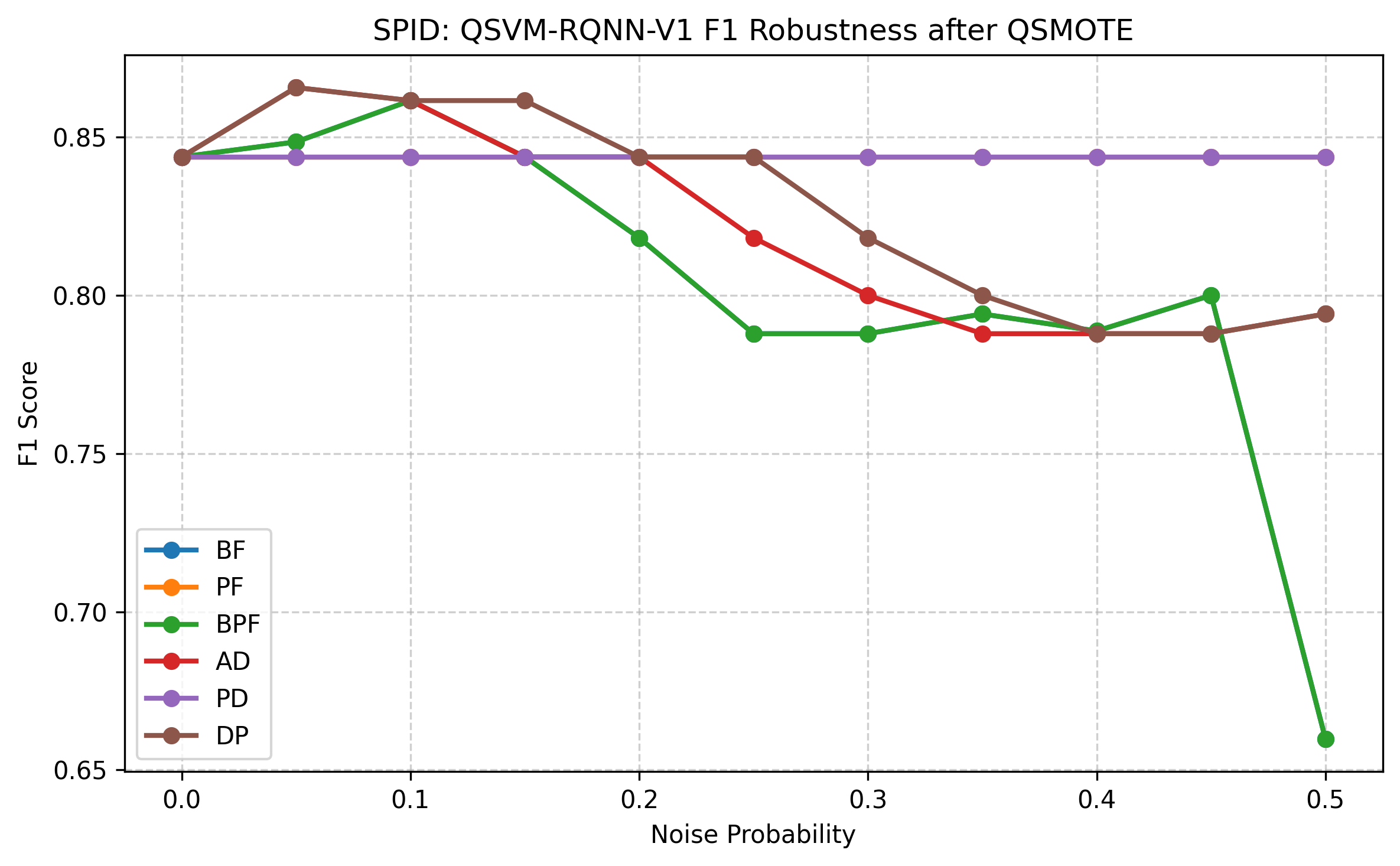}
    \caption{QSVM-RQNN-V1 with raw encoding.}
    \label{fig:spid_noise_v1}
\end{subfigure}
\hfill
\begin{subfigure}{0.48\textwidth}
    \centering
    \includegraphics[width=\linewidth]{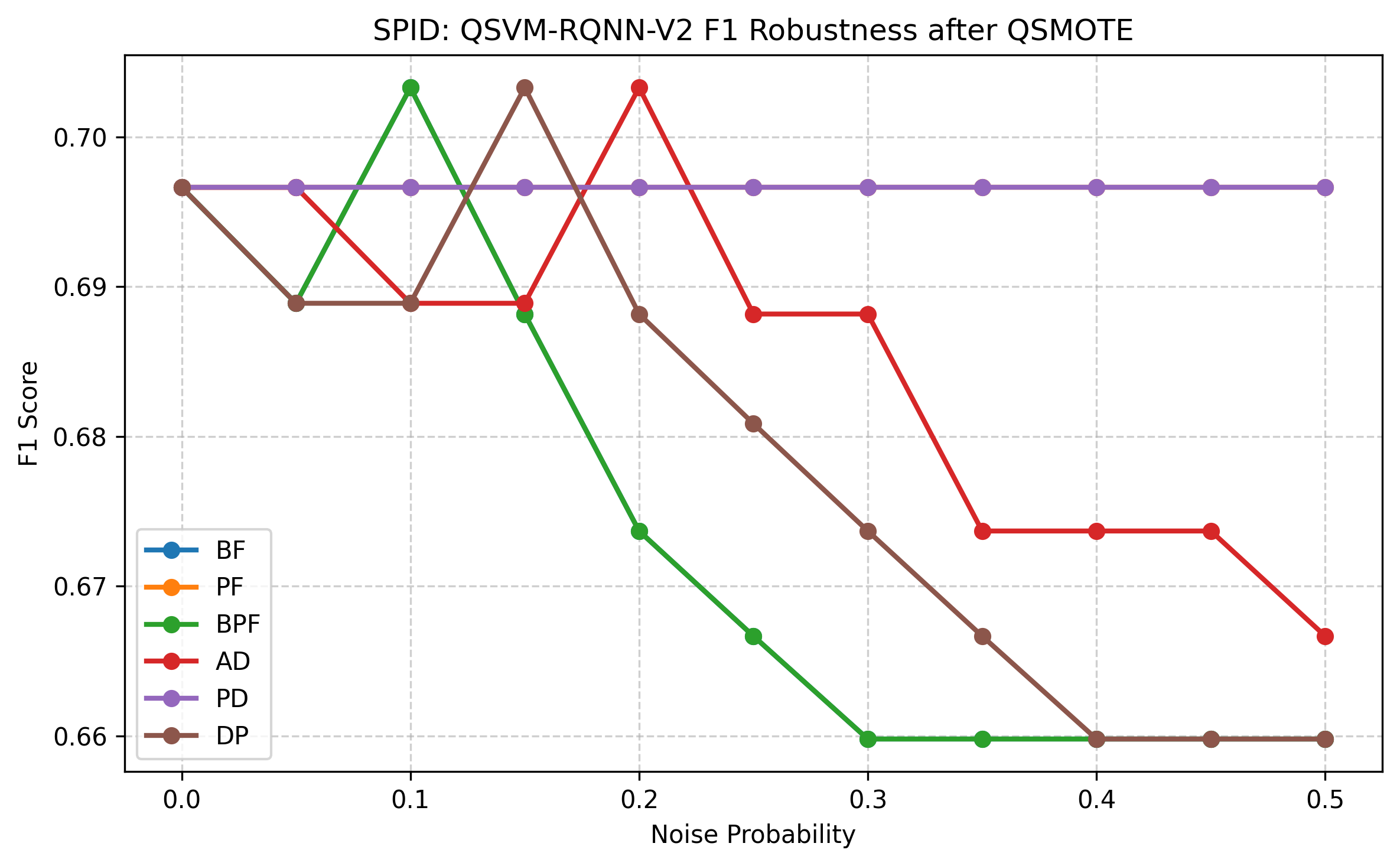}
    \caption{QSVM-RQNN-V2 with arccos encoding.}
    \label{fig:spid_noise_v2}
\end{subfigure}

\caption{Noise robustness analysis of the proposed QSVM-RQNN variants on the SPID dataset after QSMOTE. The F1-score is evaluated under six quantum noise channels, namely bit-flip (BF), phase-flip (PF), bit-phase-flip (BPF), amplitude damping (AD), phase damping (PD), and depolarizing (DP), with the noise probability varied from $\eta=0$ to $\eta=0.5$. QSVM-RQNN-V1 consistently exhibits higher robustness than QSVM-RQNN-V2.}
\label{fig:spid_noise_qsvm_rqnn}
\end{figure}

Figure~\ref{fig:spid_noise_qsvm_rqnn} compares the F1-score degradation of QSVM-RQNN-V1 and QSVM-RQNN-V2 under different quantum noise channels on the SPID dataset after QSMOTE. For QSVM-RQNN-V1, the F1-score remains relatively high across the considered noise interval, indicating strong robustness under moderate noise. The phase-damping channel shows almost no degradation, while bit-flip and phase-flip channels overlap with other curves, suggesting that their effect on the final decision probability is either identical or negligible for the learned circuit. The bit-phase-flip and depolarizing channels introduce more visible performance variation, with bit-phase-flip producing the sharpest degradation near higher noise probabilities. In contrast, QSVM-RQNN-V2 exhibits a lower overall F1-score range and stronger sensitivity to noise, particularly under bit-phase-flip, amplitude-damping, and depolarizing channels. Although phase damping remains nearly stable for V2 as well, the overall degradation pattern shows that V1 is more robust than V2 under noisy inference. This supports the selection of QSVM-RQNN-V1 as the representative architecture for the subsequent robustness analysis.

\begin{figure}[pos=htbp]
\centering

\begin{subfigure}{0.48\textwidth}
    \centering
    \includegraphics[width=\linewidth]{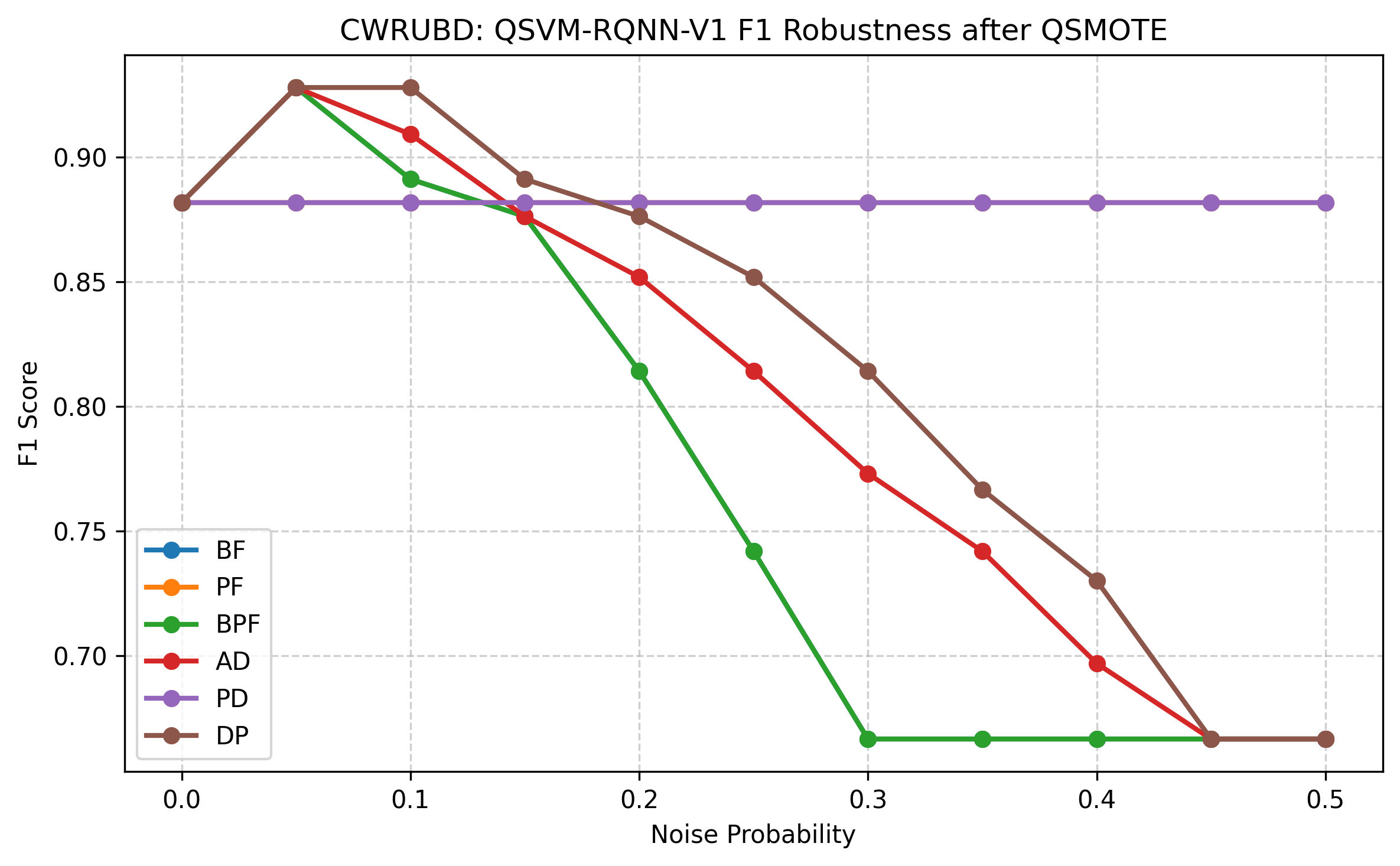}
    \caption{QSVM-RQNN-V1 with arctan encoding.}
    \label{fig:cwrubd_noise_v1}
\end{subfigure}
\hfill
\begin{subfigure}{0.48\textwidth}
    \centering
    \includegraphics[width=\linewidth]{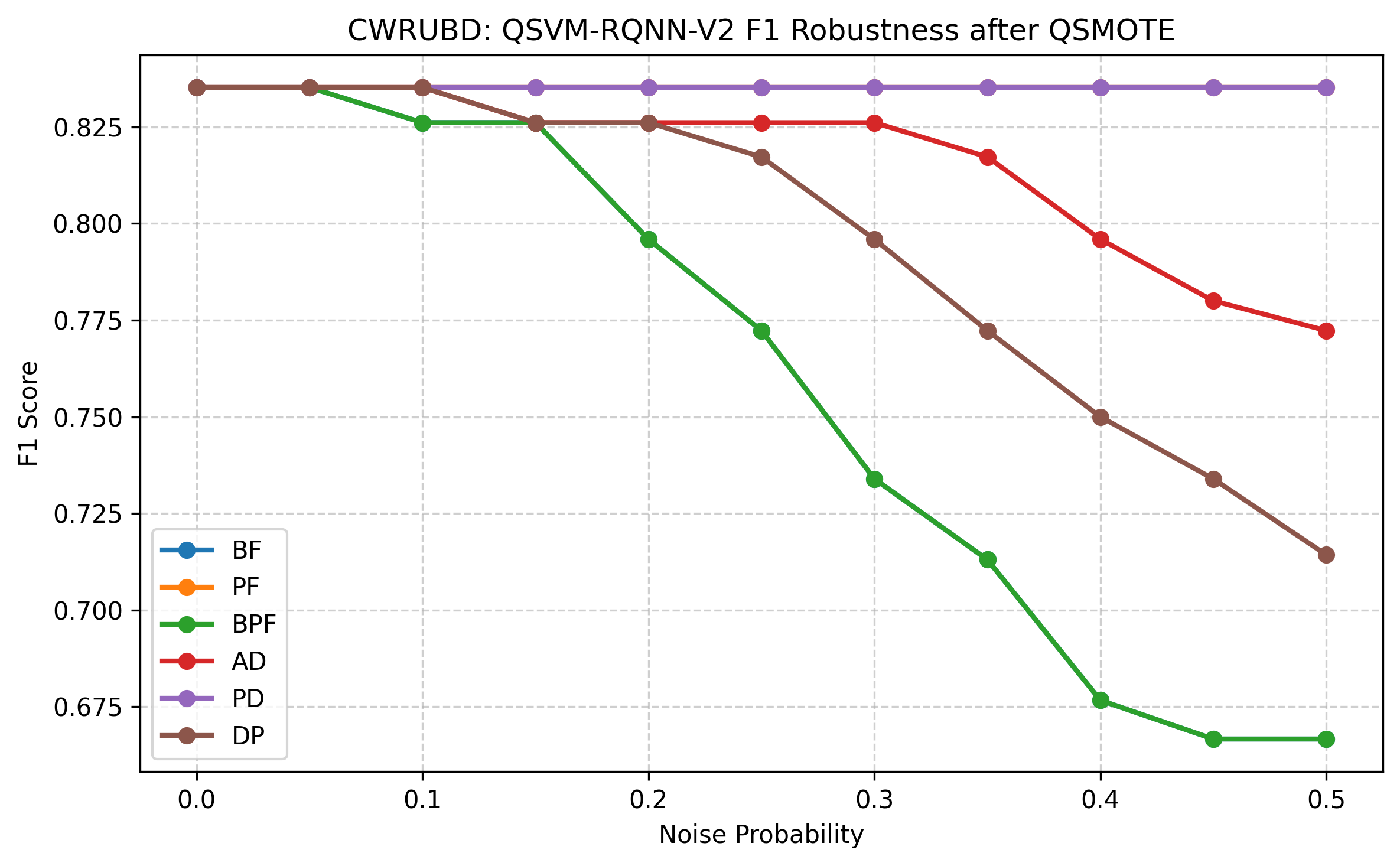}
    \caption{QSVM-RQNN-V2 with arccos encoding.}
    \label{fig:cwrubd_noise_v2}
\end{subfigure}

\caption{Noise robustness analysis of the proposed QSVM-RQNN variants on the CWRUBD dataset after QSMOTE. The F1-score is evaluated under six quantum noise channels, namely bit-flip (BF), phase-flip (PF), bit-phase-flip (BPF), amplitude damping (AD), phase damping (PD), and depolarizing (DP), while the noise probability is varied from $\eta=0$ to $\eta=0.5$. Phase damping produces almost no degradation.}
\label{fig:cwrubd_noise_qsvm_rqnn}
\end{figure}

Figure~\ref{fig:cwrubd_noise_qsvm_rqnn} presents the robustness of the proposed QSVM-RQNN-V1 and QSVM-RQNN-V2 models under different quantum noise channels on the CWRUBD dataset after QSMOTE. Both variants exhibit strong resilience against phase damping, with the F1-score remaining nearly constant throughout the investigated noise range, indicating that the learned quantum representations are largely insensitive to dephasing errors. In contrast, bit-phase-flip, amplitude-damping, and depolarizing noise gradually reduce the classification performance as the noise probability increases, with bit-phase-flip producing the fastest degradation in both variants. Comparing the two architectures, QSVM-RQNN-V1 consistently maintains higher F1-scores across almost the entire noise range and exhibits a slower degradation under amplitude-damping and depolarizing channels, demonstrating superior robustness. QSVM-RQNN-V2 also remains stable under moderate noise but shows comparatively larger performance degradation at higher noise probabilities. These observations indicate that the proposed QSVM-RQNN framework is robust against realistic quantum noise, while QSVM-RQNN-V1 provides the strongest overall noise resilience on the CWRUBD dataset.

\begin{figure}[pos=htbp]
\centering
\begin{subfigure}[b]{0.49\textwidth}
    \centering
    \includegraphics[width=\textwidth]{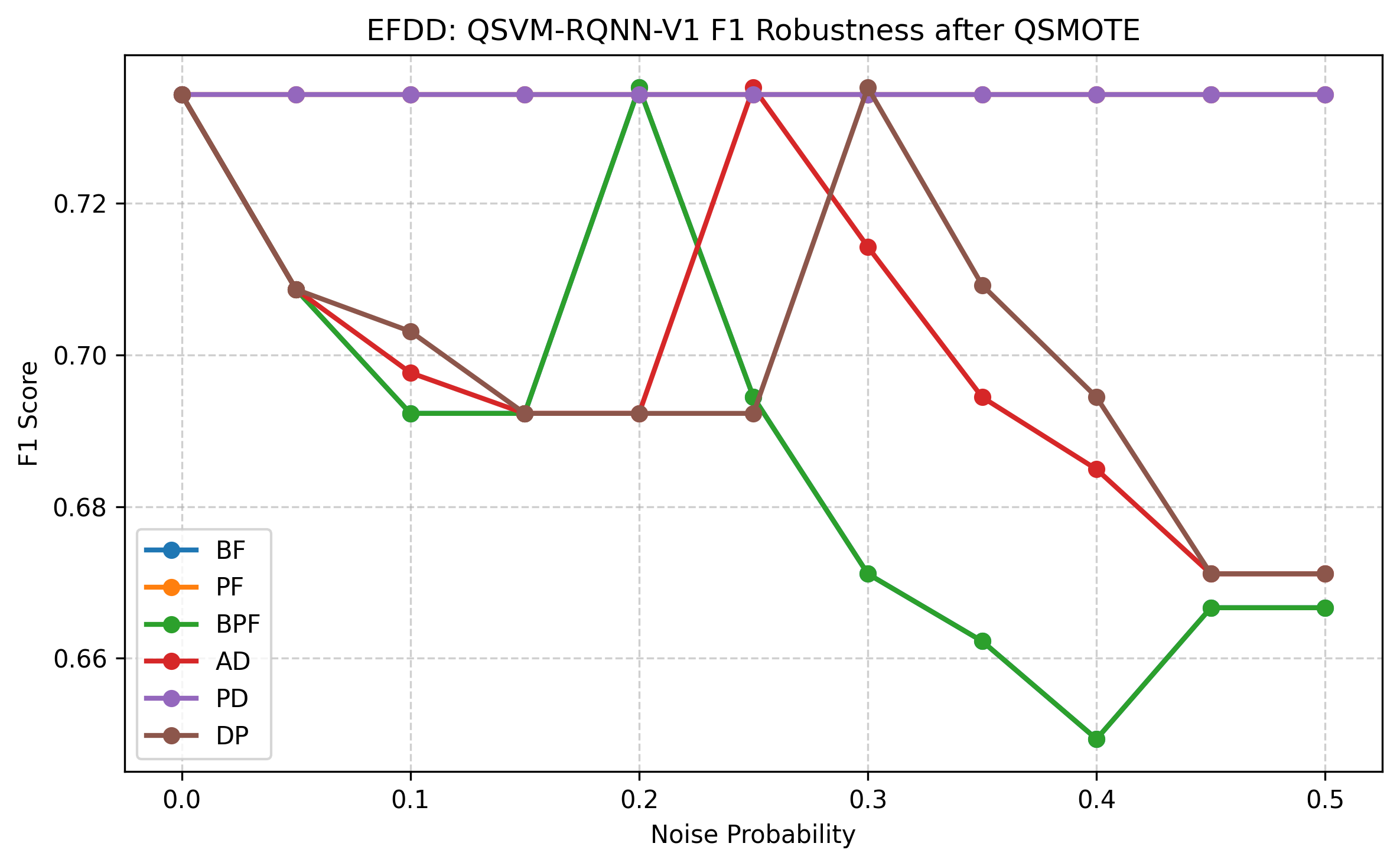}
    \caption{QSVM-RQNN-V1}
    \label{fig:efdd_noise_v1}
\end{subfigure}
\hfill
\begin{subfigure}[b]{0.49\textwidth}
    \centering
    \includegraphics[width=\textwidth]{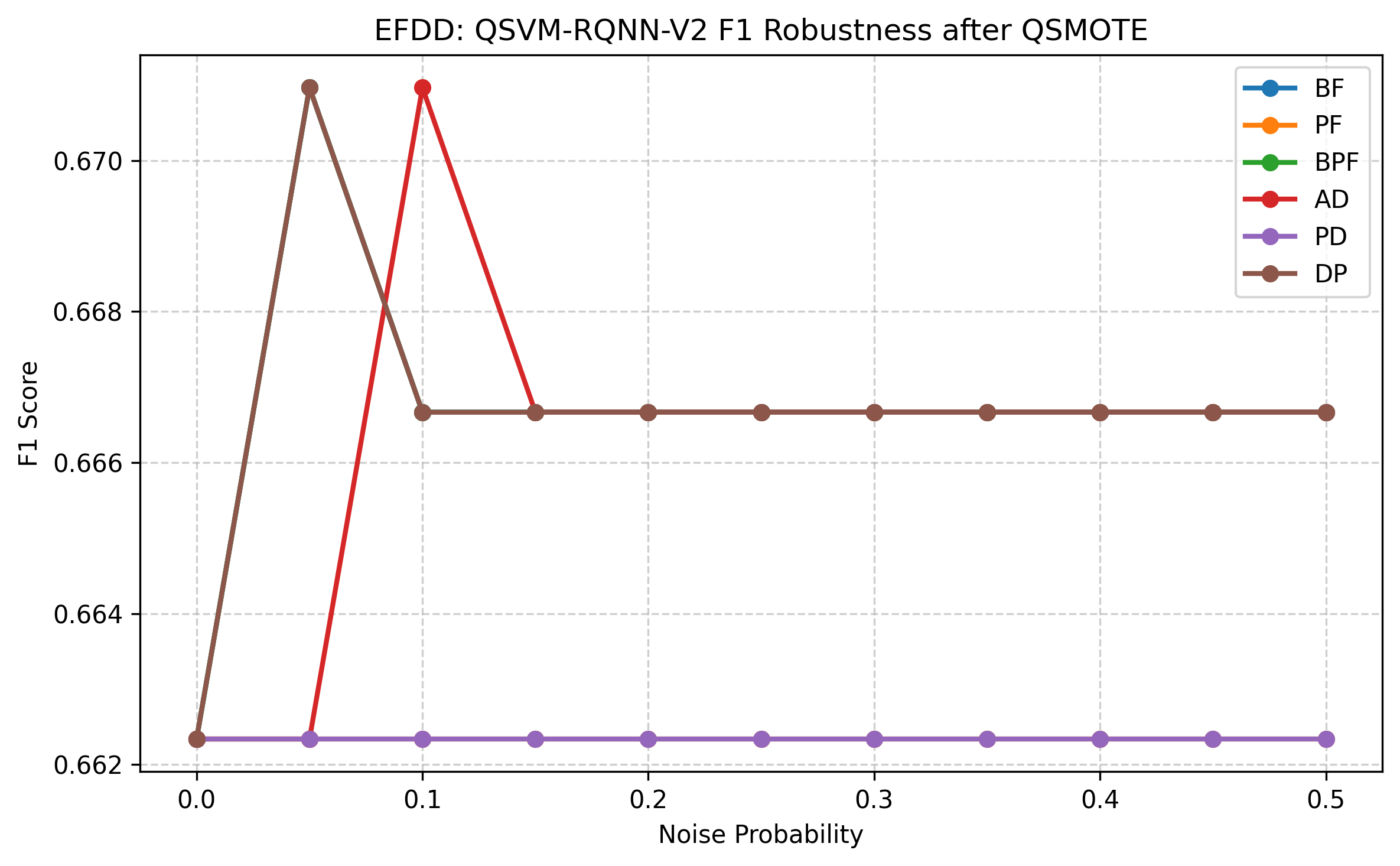}
    \caption{QSVM-RQNN-V2}
    \label{fig:efdd_noise_v2}
\end{subfigure}

\caption{Robustness analysis of the proposed QSVM-RQNN architectures on the EFDD dataset after QSMOTE under six quantum noise models: Bit Flip (BF), Phase Flip (PF), Bit-Phase Flip (BPF), Amplitude Damping (AD), Phase Damping (PD), and Depolarizing (DP). The F1-score is reported for noise probabilities ranging from $\eta=0$ to $\eta=0.5$. Both variants exhibit remarkable robustness.}
\label{fig:efdd_noise}
\end{figure}

Figure~\ref{fig:efdd_noise} presents the robustness of the proposed QSVM-RQNN architectures on the EFDD dataset after QSMOTE under six representative quantum noise channels. Compared with the previous datasets, the EFDD dataset exhibits noticeably stronger robustness to quantum noise. As shown in Fig.~\ref{fig:efdd_noise_v1}, QSVM-RQNN-V1 follows the expected degradation trend under Bit-Phase Flip (BPF), Amplitude Damping (AD), and Depolarizing (DP) noise, while Phase Flip (PF) and Phase Damping (PD) produce almost no degradation throughout the investigated noise range. The overall decrease in F1-score remains gradual, indicating that the recurrent similarity-based architecture retains reasonable classification capability even under moderate noise. An even stronger robustness is observed for QSVM-RQNN-V2 in Fig.~\ref{fig:efdd_noise_v2}. The F1-score remains nearly constant under BF, PF, and PD channels, with only very small fluctuations under AD and DP noise at low noise probabilities. These slight variations originate from minor changes in the decision boundary around the classification threshold rather than genuine performance improvements, which is expected for highly imbalanced datasets such as EFDD. Both proposed architectures demonstrate stable behavior under realistic quantum noise, with QSVM-RQNN-V2 exhibiting the highest robustness on the EFDD dataset, suggesting that its two-output recurrent decision mechanism provides improved resilience against quantum hardware imperfections.

\begin{figure}[pos=htbp]
\centering

\begin{subfigure}{0.48\textwidth}
    \centering
    \includegraphics[width=\linewidth]{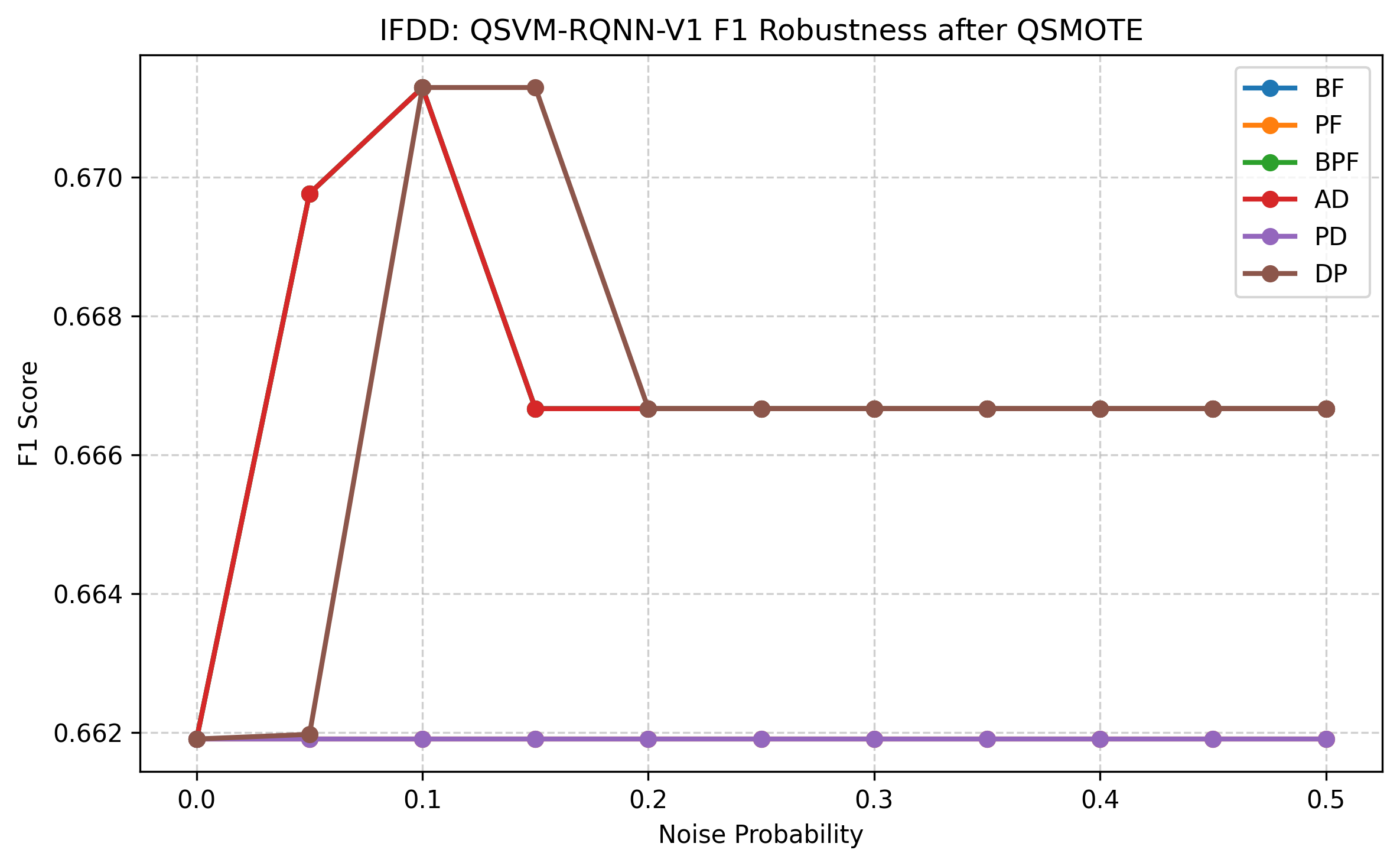}
    \caption{QSVM-RQNN-V1 with arctan encoding.}
    \label{fig:ifdd_noise_v1}
\end{subfigure}
\hfill
\begin{subfigure}{0.48\textwidth}
    \centering
    \includegraphics[width=\linewidth]{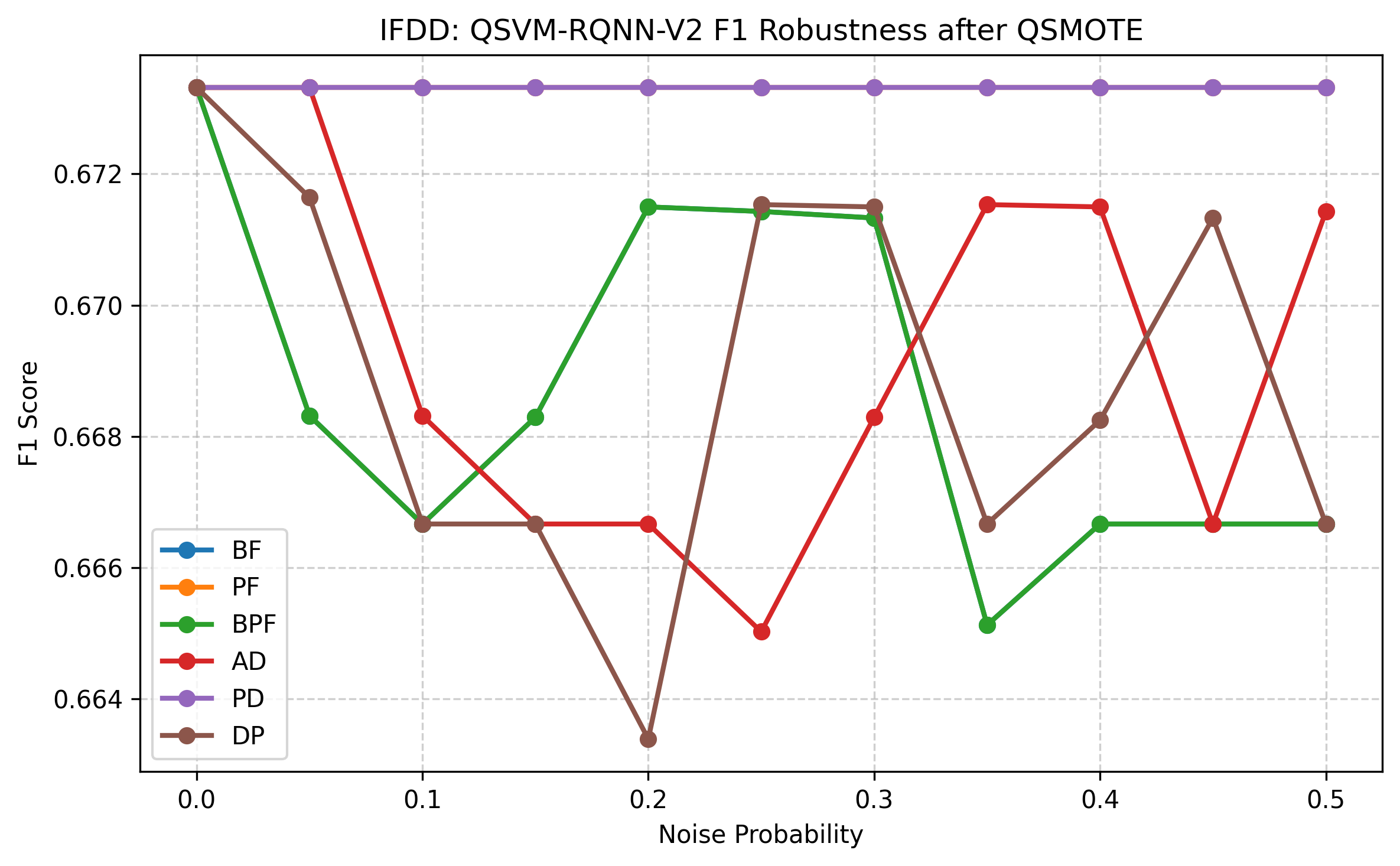}
    \caption{QSVM-RQNN-V2 with arcsin encoding.}
    \label{fig:ifdd_noise_v2}
\end{subfigure}

\caption{Noise robustness analysis of the proposed QSVM-RQNN variants on the IFDD dataset after QSMOTE. The F1-score is evaluated under six quantum noise channels: BF, PF, BPF, AD, PD, and DP, with noise probability varied from $\eta=0$ to $\eta=0.5$. Stable performance is preserved even under moderate noise.}
\label{fig:ifdd_noise_qsvm_rqnn}
\end{figure}

Figure~\ref{fig:ifdd_noise_qsvm_rqnn} shows the noise robustness behavior of QSVM-RQNN-V1 and QSVM-RQNN-V2 on the IFDD dataset after QSMOTE. Both variants exhibit highly stable F1-scores across the considered noise range, with only small fluctuations under bit flip, amplitude-damping, and depolarizing noise. Phase-damping and phase flip remain nearly invariant for both architectures, indicating no influence on the final F1-score. For QSVM-RQNN-V1, minor improvements are observed at low noise probabilities under AD and DP, which can be attributed to threshold-level perturbations rather than genuine noise-assisted learning. QSVM-RQNN-V2 also shows small oscillations under BPF, AD, and DP, but the overall F1-score remains within a narrow range. These results indicate that the proposed QSVM-RQNN framework maintains stable fault-classification performance on IFDD under moderate NISQ noise conditions.

\subsection{Ablation Study}
\begin{table}[pos=htbp]
\centering
\caption{Ablation study of QSVM-RQNN-V1 by varying the number of timesteps ($T$) while fixing the recurrent depth at $L=2$. The F1-score is reported for each dataset after QSMOTE preprocessing.}
\label{tab:ablation_v1_timesteps}
\begin{tabular}{lcccc}
\toprule
\textbf{Dataset} & $\mathbf{T=1}$ & $\mathbf{T=2}$ & $\mathbf{T=4}$ & $\mathbf{T=8}$ \\
\midrule
SPID   & 0.7529 & \textbf{0.8000} & 0.7561 & 0.7778 \\
CWRUBD & \textbf{0.9677} & 0.9663 & 0.9670 & 0.9474 \\
EFDD   & 0.6892 & \textbf{0.7050} & 0.6423 & 0.6797 \\
IFDD   & 0.6542 & 0.6432 & \textbf{0.6761} & 0.6540 \\
\bottomrule
\end{tabular}
\end{table}

\begin{table}[pos=htbp]
\centering
\caption{Ablation study of QSVM-RQNN-V1 by varying the recurrent depth ($L$) while fixing the number of timesteps at $T=4$. The F1-score is reported for each dataset after QSMOTE preprocessing.}
\label{tab:ablation_v1_layers}
\begin{tabular}{lccc}
\toprule
\textbf{Dataset} & $\mathbf{L=1}$ & $\mathbf{L=2}$ & $\mathbf{L=3}$ \\
\midrule
SPID   & \textbf{0.7945} & 0.7561 & 0.7778 \\
CWRUBD & 0.9583 & \textbf{0.9670} & 0.8723 \\
EFDD   & \textbf{0.7133} & 0.6423 & 0.6618 \\
IFDD   & 0.6683 & \textbf{0.6761} & 0.6380 \\
\bottomrule
\end{tabular}
\end{table}


An ablation study is conducted to investigate the influence of the temporal length ($T$) and recurrent depth ($L$) on the performance of the proposed QSVM-RQNN-V1 architecture. It should be noted that the QSMOTE procedure is stochastic and generates a different set of synthetic minority samples for each independent execution. Consequently, the ablation experiments were conducted on independently generated QSMOTE-balanced datasets, resulting in slight variations between the performance obtained at the default configuration (\(T=4, L=2\)) and the corresponding values reported in Tables. These minor differences originate from the stochastic nature of the oversampling process and the subsequent optimization, rather than from the architectural modifications introduced in the ablation study. Table~\ref{tab:ablation_v1_timesteps} reports the F1-score obtained by varying the number of timesteps while keeping the recurrent depth fixed at $L=2$. The results indicate that the optimal temporal resolution is dataset dependent. SPID and EFDD achieve their best performance at $T=2$, CWRUBD performs best with a single timestep ($T=1$), whereas IFDD benefits from a longer temporal representation with $T=4$. Increasing the temporal length beyond the optimal value does not consistently improve performance, suggesting that excessively long sequences may introduce redundant temporal information.

Table~\ref{tab:ablation_v1_layers} evaluates the effect of the recurrent depth while fixing $T=4$. For SPID and EFDD, a shallow recurrent architecture ($L=1$) provides the highest F1-score, whereas CWRUBD and IFDD achieve their best performance using two recurrent layers ($L=2$). Increasing the recurrent depth to three layers consistently reduces performance across all datasets, indicating that deeper recurrent quantum circuits may suffer from increased optimization difficulty without providing additional discriminative capability. The ablation study demonstrates that both the temporal length and recurrent depth substantially influence the predictive performance of QSVM-RQNN-V1, and moderate architectural configurations generally provide the best balance between representational capacity and optimization stability.

\begin{table}[pos=htbp]
\centering
\caption{Ablation study of QSVM-RQNN-V2 by varying the number of recurrent timesteps ($T$) while fixing the recurrent depth at $L=2$. The reported values correspond to the F1-score after QSMOTE preprocessing.}
\label{tab:v2_ablation_T}
\begin{tabular}{lcccc}
\toprule
\textbf{Dataset} & $\mathbf{T=1}$ & $\mathbf{T=2}$ & $\mathbf{T=4}$ & $\mathbf{T=8}$ \\
\midrule
SPID    & 0.6970 & 0.6452 & \textbf{0.7000} & 0.6598 \\
CWRUBD  & 0.7080 & 0.8250 & \textbf{0.9053} & 0.6667 \\
EFDD    & \textbf{0.6667} & 0.6667 & 0.6667 & 0.6667 \\
IFDD    & 0.6667 & \textbf{0.6745} & 0.6471 & 0.6667 \\
\bottomrule
\end{tabular}
\end{table}

\begin{table}[pos=htbp]
\centering
\caption{Ablation study of QSVM-RQNN-V2 by varying the recurrent depth ($L$) while fixing the number of timesteps at $T=4$. The reported values correspond to the F1-score after QSMOTE preprocessing.}
\label{tab:v2_ablation_L}
\begin{tabular}{lccc}
\toprule
\textbf{Dataset} & $\mathbf{L=1}$ & $\mathbf{L=2}$ & $\mathbf{L=3}$ \\
\midrule
SPID    & \textbf{0.7636} & 0.7000 & 0.6667 \\
CWRUBD  & 0.8842 & \textbf{0.9053} & 0.7143 \\
EFDD    & 0.6581 & 0.6667 & \textbf{0.6710} \\
IFDD    & 0.6495 & 0.6471 & \textbf{0.6744} \\
\bottomrule
\end{tabular}
\end{table}

Tables~\ref{tab:v2_ablation_T} and~\ref{tab:v2_ablation_L} present the ablation analysis of QSVM-RQNN-V2 by investigating the influence of the number of recurrent timesteps ($T$) and recurrent depth ($L$), respectively. For the SPID and CWRUBD datasets, increasing the temporal modeling capability up to $T=4$ yields the best performance, indicating that moderate temporal decomposition effectively captures the underlying data characteristics. However, a further increase to $T=8$ degrades the performance, suggesting that excessive sequence fragmentation weakens the discriminative information available to the recurrent quantum model. For the EFDD dataset, the F1-score remains nearly unchanged across different values of $T$, indicating that the dataset exhibits limited temporal dependence and is comparatively insensitive to sequence partitioning. In contrast, the IFDD dataset achieves its highest performance at $T=2$, while larger values of $T$ provide no additional benefit.

The influence of the recurrent depth is dataset dependent. The SPID dataset achieves the highest F1-score using a single recurrent layer ($L=1$), whereas CWRUBD benefits from a moderate depth ($L=2$), producing the best overall performance. For EFDD and IFDD, a slightly deeper recurrent architecture ($L=3$) provides marginal improvements over shallower configurations. These results demonstrate that the temporal modeling capability and recurrent depth should be selected according to the complexity of the underlying dataset. Excessively large values of either $T$ or $L$ do not necessarily improve performance and may instead introduce redundant quantum operations without additional discriminative power.

\subsection{Statistical Tests}

\begin{table}[pos=htbp]
\centering
\caption{Statistical comparison of the quantum and hybrid quantum models using post-QSMOTE F1-scores across the four benchmark datasets. Lower average rank indicates better overall performance.}
\label{tab:statistical_analysis}
\begin{tabular}{lc}
\toprule
\textbf{Model} & \textbf{Average Rank} \\
\midrule
QSVM-RQNN-V1 & \textbf{2.750} \\
RQNN-V1 & 2.875 \\
QSVM-QNN & 3.250 \\
RQNN-V2 & 4.750 \\
QSVM-RQNN-V2 & 5.375 \\
QSVM & 5.750 \\
QNN & 6.375 \\
QSVM-QCNN & 6.750 \\
QCNN & 7.125 \\
\midrule
\multicolumn{2}{c}{\textbf{Friedman Test}}\\
\midrule
Statistic & 12.303 \\
$p$-value & 0.1382 \\
\bottomrule
\end{tabular}
\end{table}

To further assess the consistency of the competing quantum models across multiple datasets, a non-parametric statistical analysis was conducted using the post-QSMOTE F1-scores. Since multiple learning algorithms were evaluated over four benchmark datasets, the Friedman test was employed to compare their overall rankings without assuming normality. Table~\ref{tab:statistical_analysis} summarizes the average ranks together with the Friedman test statistics. Among all competing approaches, QSVM-RQNN-V1 achieved the best overall average rank of 2.750, followed closely by RQNN-V1 (2.875) and QSVM-QNN (3.250), indicating that the proposed architecture consistently remained among the top-performing models across different datasets. QSVM-RQNN-V2 also demonstrated competitive performance with an average rank of 5.375, outperforming several existing quantum and hybrid quantum architectures. Although the Friedman test yielded a statistic of 12.303 with a corresponding $p$-value of 0.1382, which does not reach the conventional 0.05 significance level, this outcome is expected because only four benchmark datasets were available for statistical comparison, resulting in limited statistical power. Nevertheless, the ranking analysis consistently places QSVM-RQNN-V1 as the highest-ranked model, supporting the robustness and competitive generalization ability of the proposed architecture across diverse fault diagnosis datasets.

\subsection{Discussion}
The proposed QSVM-RQNN architectures were evaluated from multiple complementary perspectives, including predictive performance, minority-fault detection, robustness against quantum noise, architectural complexity, resource efficiency, optimization behavior, and statistical significance. Rather than relying solely on overall classification accuracy, the experimental analysis provides a comprehensive assessment of the practical characteristics that determine the suitability of quantum machine learning models for industrial fault diagnosis. Across the four benchmark datasets, the proposed architectures consistently demonstrated competitive and, in several cases, superior performance compared with the investigated quantum and hybrid quantum baselines. In particular, QSVM-RQNN-V1 achieved the best overall average rank in the Friedman analysis, while both proposed architectures maintained competitive performance across heterogeneous datasets with substantially different sensing modalities, including solar-panel inspection, bearing vibration analysis, engine telemetry, and industrial IoT monitoring.

The architectural analysis reveals that the observed performance improvements are not solely attributable to increasing the number of trainable parameters. Instead, they originate from the integration of QSVM-based similarity estimation with recurrent quantum feature refinement. While conventional quantum neural networks directly optimize a classification boundary, the proposed framework first evaluates similarity with representative class centroids before progressively refining temporal feature representations through recurrent quantum processing. This complementary interaction enables improved discrimination of minority fault samples while maintaining a relatively compact quantum circuit. The ablation study further validates the contribution of the recurrent architecture. Varying the number of recurrent timesteps and recurrent blocks demonstrates that both temporal depth and recurrent feature refinement contribute to the overall predictive performance. Increasing the temporal sequence length generally improves the ability to capture sequential feature dependencies until an appropriate operating point is reached, whereas excessively long sequences provide limited additional benefit because the available feature information has already been sufficiently represented. Similarly, increasing the recurrent depth improves representation learning up to a moderate number of recurrent blocks, after which performance tends to stabilize. These observations indicate that the selected architectural configuration provides a favorable balance between representational capability and circuit complexity.

The quantum noise experiments demonstrate that both QSVM-RQNN variants preserve stable predictive performance under realistic quantum noise channels. Although all noise models gradually degrade performance as the noise probability increases, the degradation remains smooth without abrupt instability, indicating that the proposed recurrent similarity-based framework exhibits satisfactory robustness against common quantum hardware imperfections. Furthermore, the different noise channels affect the models differently, with phase-related channels generally introducing smaller degradation than depolarizing and amplitude damping processes, which is consistent with their physical influence on quantum state evolution. Resource-efficiency analysis provides additional practical insight into the proposed architectures. Although several baseline models achieve competitive predictive performance, the proposed QSVM-RQNN architectures maintain high parameter efficiency while operating with only three qubits and a relatively small number of trainable parameters and entangling gates. The efficiency metrics demonstrate that the proposed models achieve favorable predictive performance per trainable parameter and per qubit, indicating that the observed improvements are obtained through architectural design rather than increased computational resources.

It should also be noted that the robustness analysis and ablation experiments were conducted using independently generated QSMOTE-balanced datasets. Since QSMOTE is inherently stochastic, independently generated synthetic minority samples may produce slight differences in the baseline performance observed across different experimental sections. These variations are expected and arise from the stochastic oversampling process rather than from the proposed quantum architectures or the investigated noise models. The present study focuses on binary fault-classification tasks derived from multiclass industrial benchmark datasets in order to consistently evaluate minority-fault detection, false-negative reduction, and QSMOTE-based balancing across diverse application domains. Although this experimental setting enables a controlled comparison among existing quantum models, future research may extend the proposed QSVM-RQNN framework to native multiclass quantum classification, larger-scale quantum feature spaces, adaptive quantum similarity learning, and implementation on real noisy quantum hardware.

\section{Conclusion}\label{SecV}

This paper presented QSVM-RQNN, a hybrid quantum machine learning framework that integrates quantum support vector machine (QSVM) similarity estimation with recurrent quantum neural network (RQNN) learning for fault diagnosis in industrial systems. Two complementary architectures, namely QSVM-RQNN-V1 and QSVM-RQNN-V2, were developed to investigate similarity-based and dual-output recurrent quantum decision mechanisms while maintaining a compact three-qubit implementation. To address the class imbalance commonly encountered in industrial fault diagnosis, a QSMOTE was introduced to generate balanced quantum feature representations prior to model training. The proposed framework was evaluated on binary fault-classification tasks derived from four widely used benchmark datasets representing solar-panel inspection, bearing fault diagnosis, engine failure detection, and industrial IoT monitoring.

Comprehensive experimental evaluation demonstrated that the proposed architectures achieve competitive and, in several cases, superior performance compared with existing quantum and hybrid quantum models. In addition to predictive performance, the proposed framework was systematically analyzed through architectural complexity evaluation, resource-efficiency analysis, quantum noise robustness assessment, ablation studies, and statistical significance analysis. The results indicate that combining QSVM-based similarity estimation with recurrent quantum feature refinement enables improved minority-fault detection while maintaining low quantum resource requirements. Furthermore, the proposed architectures exhibited stable behavior under multiple realistic quantum noise models, highlighting their potential suitability for near-term quantum computing platforms. Although the present work focuses on binary fault-classification tasks derived from multiclass benchmark datasets, the proposed framework provides a flexible foundation for more general quantum fault diagnosis systems. Future research will investigate native multiclass quantum classification, adaptive quantum similarity learning, deeper recurrent quantum architectures, real quantum hardware implementation, and the integration of advanced quantum optimization strategies for large-scale industrial monitoring applications.

\input{arXiv.bbl}

\end{document}

%% file: arXiv.bbl

%% file: arXiv.bbl
\begin{thebibliography}{10}
\providecommand{\url}[1]{#1}
\csname url@samestyle\endcsname
\providecommand{\newblock}{\relax}
\providecommand{\bibinfo}[2]{#2}
\providecommand{\BIBentrySTDinterwordspacing}{\spaceskip=0pt\relax}
\providecommand{\BIBentryALTinterwordstretchfactor}{4}
\providecommand{\BIBentryALTinterwordspacing}{\spaceskip=\fontdimen2\font plus
\BIBentryALTinterwordstretchfactor\fontdimen3\font minus \fontdimen4\font\relax}
\providecommand{\BIBforeignlanguage}[2]{{%
\expandafter\ifx\csname l@#1\endcsname\relax
\typeout{** WARNING: IEEEtran.bst: No hyphenation pattern has been}%
\typeout{** loaded for the language `#1'. Using the pattern for}%
\typeout{** the default language instead.}%
\else
\language=\csname l@#1\endcsname
\fi
#2}}
\providecommand{\BIBdecl}{\relax}
\BIBdecl

\bibitem{lee2015cyber}
\BIBentryALTinterwordspacing
J.~Lee, B.~Bagheri, and H.-A. Kao, ``A cyber-physical systems architecture for industry 4.0-based manufacturing systems,'' \emph{Manufacturing Letters}, vol.~3, pp. 18--23, 2015. [Online]. Available: \url{https://doi.org/10.1016/j.mfglet.2014.12.001}
\BIBentrySTDinterwordspacing

\bibitem{zhang2019deep}
\BIBentryALTinterwordspacing
L.~Wu and X.~Sun, ``A review of deep learning-based fault diagnosis techniques for axlebox bearings in rail vehicles,'' in \emph{2025 IEEE International Conference on Intelligent Rail Transportation (ICIRT)}.\hskip 1em plus 0.5em minus 0.4em\relax Beijing, China: IEEE, 2025, conference held October 11--12, 2025. [Online]. Available: \url{https://ieeexplore.ieee.org/document/11216733}
\BIBentrySTDinterwordspacing

\bibitem{yan2020industrial}
\BIBentryALTinterwordspacing
J.~Yan, Y.~Meng, L.~Lu, and L.~Li, ``Industrial big data in an industry 4.0 environment: Challenges, schemes, and applications for predictive maintenance,'' \emph{IEEE Access}, vol.~5, pp. 23\,484--23\,491, 2017. [Online]. Available: \url{https://ieeexplore.ieee.org/document/8085101}
\BIBentrySTDinterwordspacing

\bibitem{lecun2015deep}
\BIBentryALTinterwordspacing
Y.~LeCun, Y.~Bengio, and G.~Hinton, ``Deep learning,'' \emph{Nature}, vol. 521, pp. 436--444, 2015. [Online]. Available: \url{https://doi.org/10.1038/nature14539}
\BIBentrySTDinterwordspacing

\bibitem{goodfellow2016deep}
\BIBentryALTinterwordspacing
I.~Goodfellow, Y.~Bengio, and A.~Courville, \emph{Deep Learning}.\hskip 1em plus 0.5em minus 0.4em\relax Cambridge, MA: MIT Press, 2016. [Online]. Available: \url{https://www.deeplearningbook.org}
\BIBentrySTDinterwordspacing

\bibitem{wang2021deep}
\BIBentryALTinterwordspacing
Y.~Wang, X.~Ge, H.~Ma, S.~Qi, G.~Zhang, and Y.~Yao, ``Deep learning in medical ultrasound image analysis: A review,'' \emph{IEEE Access}, vol.~9, pp. 54\,310--54\,324, 2021. [Online]. Available: \url{https://ieeexplore.ieee.org/document/9395635}
\BIBentrySTDinterwordspacing

\bibitem{nielsen2010quantum}
\BIBentryALTinterwordspacing
M.~A. Nielsen and I.~L. Chuang, \emph{Quantum Computation and Quantum Information}.\hskip 1em plus 0.5em minus 0.4em\relax Cambridge University Press, 2010. [Online]. Available: \url{https://doi.org/10.1017/CBO9780511976667}
\BIBentrySTDinterwordspacing

\bibitem{preskill2018quantum}
\BIBentryALTinterwordspacing
J.~Preskill, ``Quantum {C}omputing in the {NISQ} era and beyond,'' \emph{{Quantum}}, vol.~2, p.~79, Aug. 2018. [Online]. Available: \url{https://doi.org/10.22331/q-2018-08-06-79}
\BIBentrySTDinterwordspacing

\bibitem{schuld2015introduction}
\BIBentryALTinterwordspacing
M.~Schuld, I.~Sinayskiy, and F.~Petruccione, ``An introduction to quantum machine learning,'' \emph{Contemporary Physics}, vol.~56, no.~2, pp. 172--185, 2015. [Online]. Available: \url{https://doi.org/10.1080/00107514.2014.964942}
\BIBentrySTDinterwordspacing

\bibitem{biamonte2017quantum}
\BIBentryALTinterwordspacing
J.~Biamonte, P.~Wittek, N.~Pancotti, P.~Rebentrost, N.~Wiebe, and S.~Lloyd, ``Quantum machine learning,'' \emph{Nature}, vol. 549, pp. 195--202, 2017. [Online]. Available: \url{https://www.nature.com/articles/nature23474}
\BIBentrySTDinterwordspacing

\bibitem{schuld2021machine}
\BIBentryALTinterwordspacing
M.~Schuld and F.~Petruccione, \emph{Machine Learning with Quantum Computers}.\hskip 1em plus 0.5em minus 0.4em\relax Springer, 2021. [Online]. Available: \url{https://doi.org/10.1007/978-3-030-83098-4}
\BIBentrySTDinterwordspacing

\bibitem{cerezo2021variational}
\BIBentryALTinterwordspacing
M.~Cerezo, A.~Arrasmith, R.~Babbush, S.~C. Benjamin, S.~Endo, K.~Fujii, J.~R. McClean, K.~Mitarai, X.~Yuan, L.~Cincio, and P.~J. Coles, ``Variational quantum algorithms,'' \emph{Nature Reviews Physics}, vol.~3, pp. 625--644, 2021. [Online]. Available: \url{https://doi.org/10.1038/s42254-021-00348-9}
\BIBentrySTDinterwordspacing

\bibitem{farhi2018classification}
\BIBentryALTinterwordspacing
E.~Farhi and H.~Neven, ``Classification with quantum neural networks on near term processors,'' \emph{arXiv preprint arXiv:1802.06002}, 2018. [Online]. Available: \url{https://doi.org/10.48550/arXiv.1802.06002}
\BIBentrySTDinterwordspacing

\bibitem{cong2019quantum}
\BIBentryALTinterwordspacing
I.~Cong, S.~Choi, and M.~Lukin, ``Quantum convolutional neural networks,'' \emph{Nature Physics}, vol.~15, pp. 1273--1278, 2019. [Online]. Available: \url{https://doi.org/10.1038/s41567-019-0648-8}
\BIBentrySTDinterwordspacing

\bibitem{havlivcek2019supervised}
\BIBentryALTinterwordspacing
V.~Havl{\'\i}{\v{c}}ek, A.~D. C{\'o}rcoles, K.~Temme, A.~W. Harrow, A.~Kandala, J.~M. Chow, and J.~M. Gambetta, ``Supervised learning with quantum-enhanced feature spaces,'' \emph{Nature}, vol. 567, pp. 209--212, 2019. [Online]. Available: \url{https://doi.org/10.1038/s41586-019-0980-2}
\BIBentrySTDinterwordspacing

\bibitem{schuld2019quantum}
\BIBentryALTinterwordspacing
M.~Schuld and N.~Killoran, ``Quantum machine learning in feature hilbert spaces,'' \emph{Physical Review Letters}, vol. 122, no.~4, p. 040504, 2019. [Online]. Available: \url{https://doi.org/10.1103/PhysRevLett.122.040504}
\BIBentrySTDinterwordspacing

\bibitem{schuld2020circuit}
\BIBentryALTinterwordspacing
M.~Schuld, A.~Bocharov, K.~Svore, and N.~Wiebe, ``Circuit-centric quantum classifiers,'' \emph{Physical Review A}, vol. 101, p. 032308, 2020. [Online]. Available: \url{https://doi.org/10.1103/PhysRevA.101.032308}
\BIBentrySTDinterwordspacing

\bibitem{mcclean2018barren}
\BIBentryALTinterwordspacing
J.~R. McClean, S.~Boixo, R.~Smelyanskiy, R.~Babbush, and H.~Neven, ``Barren plateaus in quantum neural network training landscapes,'' \emph{Nature Communications}, vol.~9, p. 4812, 2018. [Online]. Available: \url{https://doi.org/10.1038/s41467-018-07090-4}
\BIBentrySTDinterwordspacing

\bibitem{beer2020training}
\BIBentryALTinterwordspacing
K.~Beer, D.~Bondarenko, T.~Farrelly, T.~J. Osborne, R.~Salzmann, D.~Scheiermann, and R.~Wolf, ``Training deep quantum neural networks,'' \emph{Nature Communications}, vol.~11, p. 808, 2020. [Online]. Available: \url{https://doi.org/10.1038/s41467-020-14454-2}
\BIBentrySTDinterwordspacing

\bibitem{abbas2021power}
\BIBentryALTinterwordspacing
A.~Abbas, D.~Sutter, C.~Zoufal, A.~Lucchi, A.~Figalli, and S.~Woerner, ``The power of quantum neural networks,'' \emph{Nature Computational Science}, vol.~1, pp. 403--409, 2021. [Online]. Available: \url{https://doi.org/10.1038/s43588-021-00084-1}
\BIBentrySTDinterwordspacing

\bibitem{Mohanty2025QuantumSMOTE}
\BIBentryALTinterwordspacing
N.~Mohanty, B.~K. Behera, C.~Ferrie, and P.~Dash, ``A quantum approach to synthetic minority oversampling technique (smote),'' \emph{Quantum Machine Intelligence}, vol.~7, no.~1, p.~38, 2025. [Online]. Available: \url{https://link.springer.com/article/10.1007/s42484-025-00248-6}
\BIBentrySTDinterwordspacing

\bibitem{behera2025qsvmqnn}
\BIBentryALTinterwordspacing
B.~K. Behera, S.~Al-Kuwari, and A.~Farouk, ``{QSVM-QNN}: Quantum support vector machine based quantum neural network learning algorithm for brain--computer interfacing systems,'' \emph{IEEE Transactions on Artificial Intelligence}, vol.~7, no.~1, pp. 308--320, 2025. [Online]. Available: \url{https://doi.org/10.1109/TAI.2025.3572852}
\BIBentrySTDinterwordspacing

\bibitem{cortes1995support}
\BIBentryALTinterwordspacing
C.~Cortes and V.~Vapnik, ``Support-vector networks,'' \emph{Machine Learning}, vol.~20, pp. 273--297, 1995. [Online]. Available: \url{https://doi.org/10.1007/BF00994018}
\BIBentrySTDinterwordspacing

\bibitem{vapnik1998statistical}
\BIBentryALTinterwordspacing
V.~N. Vapnik, ``An overview of statistical learning theory,'' \emph{IEEE Transactions on Neural Networks}, vol.~10, no.~5, pp. 988--999, Sep 1999. [Online]. Available: \url{https://doi.org/10.1109/72.788640}
\BIBentrySTDinterwordspacing

\bibitem{widodo2007support}
\BIBentryALTinterwordspacing
A.~Widodo and B.-S. Yang, ``Support vector machine in machine condition monitoring and fault diagnosis,'' \emph{Mechanical Systems and Signal Processing}, vol.~21, no.~6, pp. 2560--2574, 2007. [Online]. Available: \url{https://doi.org/10.1016/j.ymssp.2006.12.007}
\BIBentrySTDinterwordspacing

\bibitem{shao2011fault}
\BIBentryALTinterwordspacing
W.~Zhao, Y.~Lv, J.~Xiao, and Y.~Li, ``Fault diagnosis of rolling bearings based on {GA}-{SVM} model,'' in \emph{2021 Global Reliability and Prognostics and Health Management (PHM-Nanjing)}.\hskip 1em plus 0.5em minus 0.4em\relax Nanjing, China: IEEE, 2021, conference held October 15--17, 2021. [Online]. Available: \url{https://ieeexplore.ieee.org/document/9612886}
\BIBentrySTDinterwordspacing

\bibitem{rebentrost2014quantum}
\BIBentryALTinterwordspacing
P.~Rebentrost, M.~Mohseni, and S.~Lloyd, ``Quantum support vector machine for big data classification,'' \emph{Physical Review Letters}, vol. 113, p. 130503, 2014. [Online]. Available: \url{https://doi.org/10.1103/PhysRevLett.113.130503}
\BIBentrySTDinterwordspacing

\bibitem{li2022quantum}
\BIBentryALTinterwordspacing
Y.~Li, L.~Song, Q.~Sun, H.~Xu, X.~Li, Z.~Fang, and W.~Yao, ``Rolling bearing fault diagnosis based on quantum ls-svm,'' \emph{EPJ Quantum Technology}, vol.~9, p.~18, 2022. [Online]. Available: \url{https://doi.org/10.1140/epjqt/s40507-022-00137-y}
\BIBentrySTDinterwordspacing

\bibitem{wang2024quantum}
\BIBentryALTinterwordspacing
Q.-L. Wang, Y.~Jin, X.-H. Li, Y.~Li, Y.-C. Li, K.-J. Zhang, H.~Liu, and L.~Cheng, ``An advanced quantum support vector machine for power quality disturbance detection and identification,'' \emph{EPJ Quantum Technology}, vol.~11, p.~70, 2024. [Online]. Available: \url{https://doi.org/10.1140/epjqt/s40507-024-00283-5}
\BIBentrySTDinterwordspacing

\bibitem{diedrich2024quantum}
\BIBentryALTinterwordspacing
A.~Diedrich, S.~Windmann, and O.~Niggemann, ``Solving industrial fault diagnosis problems with quantum computers,'' \emph{Quantum Machine Intelligence}, vol.~6, p.~66, 2024. [Online]. Available: \url{https://doi.org/10.1007/s42484-024-00184-x}
\BIBentrySTDinterwordspacing

\bibitem{nakajima2021quantum}
\BIBentryALTinterwordspacing
C.~Gyurik, F.~Wudarski, E.~Philip, A.~Sannia, H.~Sadeghi, O.~Kyriienko, D.~Venturelli, and A.~A. Gentile, ``From quantum feature maps to quantum reservoir computing: Perspectives and applications,'' \emph{arXiv preprint arXiv:2510.01797}, 2025. [Online]. Available: \url{https://doi.org/10.48550/arXiv.2510.01797}
\BIBentrySTDinterwordspacing

\bibitem{fujii2023quantum}
\BIBentryALTinterwordspacing
K.~Fujii and K.~Nakajima, ``Harnessing disordered quantum dynamics for machine learning,'' \emph{Physical Review Applied}, vol.~8, no.~2, p. 024030, 2017. [Online]. Available: \url{https://doi.org/10.1103/PhysRevApplied.8.024030}
\BIBentrySTDinterwordspacing

\bibitem{afroz2023solar}
P.~Afroz, ``Solar panel images dataset,'' \url{https://www.kaggle.com/datasets/pythonafroz/solar-panel-images}, 2023, accessed: 2025-09-01.

\bibitem{cwru_kaggle}
{brjapon}, ``Cwru bearing datasets,'' \url{https://www.kaggle.com/datasets/brjapon/cwru-bearing-datasets}, kaggle dataset. Accessed: 2025-09-26.

\bibitem{engine_failure_kaggle}
{ziya07}, ``Engine failure detection dataset,'' \url{https://www.kaggle.com/datasets/ziya07/engine-failure-detection-dataset}, kaggle dataset. Accessed: 2025-09-30.

\bibitem{industrialfault2024}
Programmer3, ``Industrial fault detection dataset,'' \url{https://www.kaggle.com/datasets/programmer3/industrial-fault-detection-dataset}, 2024, accessed: October 2025.

\bibitem{mohanty2023vehicle}
\BIBentryALTinterwordspacing
N.~Mohanty, B.~K. Behera, and C.~Ferrie, ``Analysis of the vehicle routing problem solved via hybrid quantum algorithms in the presence of noisy channels,'' \emph{IEEE Transactions on Quantum Engineering}, vol.~4, pp. 1--14, 2023. [Online]. Available: \url{https://ieeexplore.ieee.org/document/10214310}
\BIBentrySTDinterwordspacing

\end{thebibliography}
